\documentclass[aip,apl,twocolumn,reprint,longbibliography]{revtex4-1}
\usepackage{graphicx}
\usepackage{dcolumn}
\usepackage{bm}
\usepackage{color}
\usepackage{tabularx}
\usepackage{array}
\usepackage{amsmath}
\usepackage{pbox}
\usepackage{stmaryrd}  
\usepackage{color,soul}

\begin{document}

\title{Coupling Magnons to an Opto-Electronic Parametric Oscillator}


\author{Junming Wu}
\affiliation{Department of Physics and Astronomy, University of North Carolina at Chapel Hill, Chapel Hill, NC 27599, USA}

\author{Shihao Zhou}
\affiliation{Department of Physics and Astronomy, University of North Carolina at Chapel Hill, Chapel Hill, NC 27599, USA}

\author{Benedetta Flebus}
\email{flebus@bc.edu} 
\affiliation{Department of Physics, Boston College, 140 Commonwealth Avenue, Chestnut Hill, MA 02467, USA}

\author{Wei Zhang}
\email{zhwei@unc.edu}
\affiliation{Department of Physics and Astronomy, University of North Carolina at Chapel Hill, Chapel Hill, NC 27599, USA}

\date{\today}

\begin{abstract}

Hybrid magnonic systems have emerged as versatile modular components for quantum signal transduction and sensing applications owing to their capability of connecting distinct quantum platforms. To date, the majority of the magnonic systems have been explored in a local, near-field scheme, due to the close proximity required for realizing a strong coupling between magnons and other excitations. This constraint greatly limits the applicability of magnons in developing remotely-coupled, distributed quantum network systems. On the contrary, opto-electronic architectures hosting self-sustained oscillations has been a unique platform for long-haul signal transmission and processing. Here, we integrated an opto-electronic oscillator with a magnonic oscillator consisting of a microwave waveguide and a $\rm Y_3Fe_5O_{12}$(YIG) sphere, and demonstrated strong and coherent coupling between YIG's magnon modes and the opto-electronic oscillator's characteristic photon modes -- revealing the hallmark anti-crossing gap in the measured spectrum. In particular, the photon mode is produced on-demand via a nonlinear, parametric process as stipulated by an external seed pump. Both the internal cavity phase and the external pump phase can be precisely tuned to stabilize either degenerate or nondegenerate auto-oscillations. Our result lays out a new, hybrid platform for investigating long-distance coupling and nonlinearity in coherent magnonic phenomena, which may be find useful in constructing future  `distributed hybrid magnonic systems'.    

\end{abstract}

\maketitle

\section{Introduction}

Hybrid magnonics have recently witnessed immense developments, with successful demonstrations of magnon hybridization with other fundamental excitations including microwave, light, phonon, and qubit, manifesting a series of coherent phenomena \cite{awschalom2021quantum,lachance2019hybrid,li2020hybrid,yuan2022quantum,flebus20242024,chumak2022advances}. Thus far, these phenomena have mostly been explored only in a `local, non-distributed' fashion dictated by the close proximity required for realizing a `strong coupling' between magnons and other excitations \cite{xiong2025photon,xiong2024hybrid,xiong2024combinatorial,xu2024slow,li2021advances}.  

Recently, the emergent developments in quantum networks and quantum interconnects \cite{wang2024nanoscale,awschalom2021development} call for the pressing need of `long-distance coupling' schemes \cite{rao2023meterscale,yang2024anomalous}, which is another important attribute in hybrid quantum systems, yet it represents a major challenge for hybrid magnonics due to the often near-field interactions involving magnons. One possible solution to this challenge is to hybridize magnonic systems with a distributed resonance architecture under an external gain \cite{yao2023coherent,zhang2025gain}. A natural candidate is the opto-electronic oscillator (OEO) \cite{chembo2019optoelectronic}, which features both high quality-factor (Q-factor) microwave photon modes (in the GHz regime) and long-haul signal transmission capability with minimum insertion loss (over kms), leveraging fiber optics and convenient Electrical(E)-Optical(O) conversion by using electro-optic modulators, and vice versa (O-E) by using ultrafast photodiodes. However, the potential of such distributed, hybrid magnonic systems have yet remained largely unexplored. 

\begin{figure*}[htb]
 \centering
 \includegraphics[width=6.9 in]{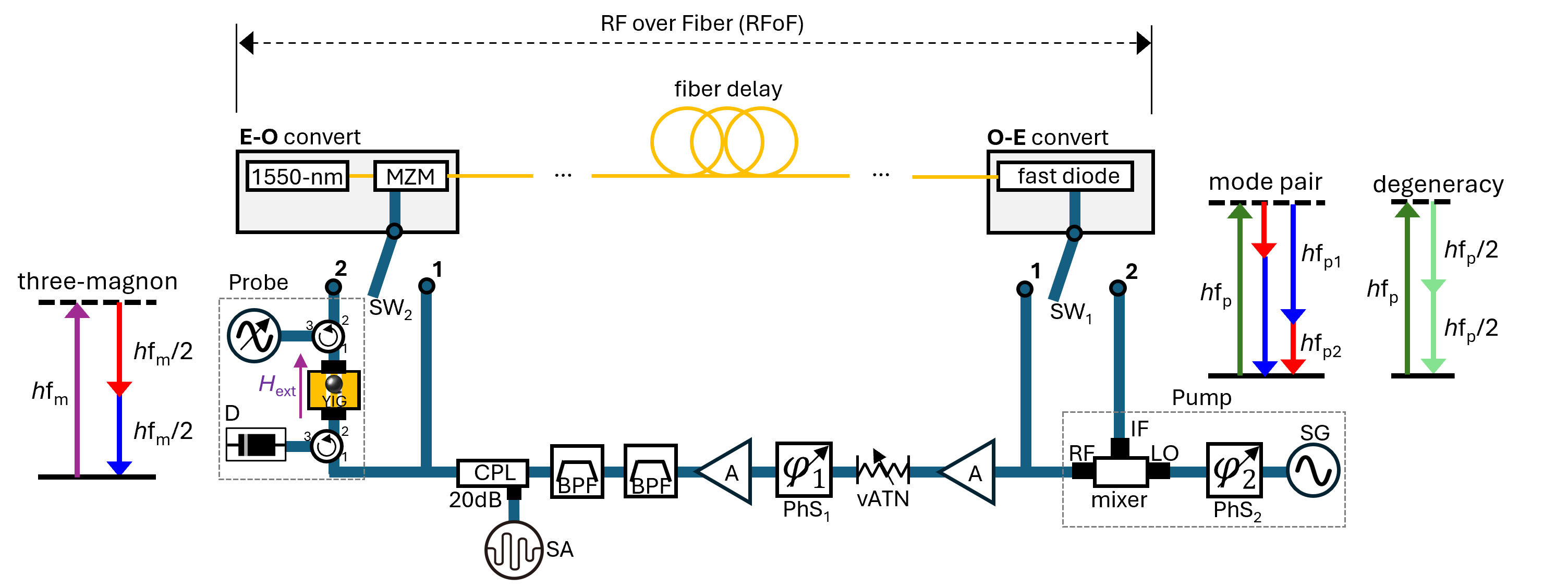}
 \caption{Schematic illustration of the OEMPO experimental setup. The central system is a standard OEO, consisting of an optical path (RFoF system with O-E and E-O conversions), and an electrical path, forming a cavity loop. An external pump (right dashline-enclosed box), injecting a pump tone, $\rm f_p$, can be integrated to the OEO and transforming the system to a standard OEPO. Either nondegenerate ($\rm f_{p1},f_{p2}$ and $\rm f_{p1}+f_{p2} = f_p$) or degenerate ($\rm \frac{f_p}{2}$) auto-oscillation modes can be induced inside the cavity. Another probe system consists of a YIG magnonic resonator (left dashline-enclosed box) can be inserted into the loop to realize coherent coupling between the magnon modes and the OEPO's photon modes, via the pump-induced three-magnon splitting and confluence process ($\rm f_m \rightarrow \frac{f_m}{2} + \frac{f_m}{2}$). SW: switch (manual connection), SG: signal generator, PhS: phase shifter, A: amplifier, vATN: variable attenuator, BPF: band-pass filter, SA: spectrum analyzer, CPL: coupler, D: diode, MZM: Mach-Zehnder modulator. The different configurations can be adjusted by the two SWs. OEO: SW$_1$ and SW$_2$ both at position-1; OEMO: SW$_1$ at position-1, SW$_2$ at position-2; OEPO: SW$_1$ at position-2, SW$_2$ at position-1. OEMPO: SW$_1$ and SW$_2$ both at position-2. }
 \label{fig:scheme}
\end{figure*}

Basically, the OEO is a type of delayline oscillator that is implemented in an optoelectronic cavity. It has a hybrid positive feedback loop formed by an optical section combined with an electrical section to create microwave signals with ultralow phase noise owing to the use of a high Q-factor optical energy-storage element, such as a long optical fiber delayline. Thus, the steady oscillation in an OEO is a delay-controlled operation, which, on the flip slide, often leads to difficulties in frequency tuning and stabilizing. The frequency tuning in a standard OEO is discrete, with the minimum tuning step determined by the cavity delay. In an prior work, the coherent coupling between magnons and OEO photons has been demonstrated \cite{xiong2024magnon}.   

On the other hand, parametric oscillator is another type of important oscillator based on a nonlinear process. A good example is the optical parametric oscillators (OPOs) \cite{dunn1999parametric}, which extend the operating frequency of lasers by utilizing 2nd- or 3rd-order nonlinearity, while the operating frequency range of ordinary lasers is limited to the stimulated atomic energies. A parametric oscillator can also be designed in the rf domain by utilizing a nonlinear electronic device. For example, an optoelectronic parametric oscillator (OEPO) was recently realized based on the 2nd-order nonlinearity in an optoelectronic cavity \cite{hao2020optoelectronic,cen2022large}, which ensures stable multimode oscillation that is difficult to realize in standard OEOs or OPOs due to the mode-hopping and mode-competition effects \cite{chembo2008effects,peil2009routes}. Owing to the unique energy-transition process in the optoelectronic cavity, oscillations in the OEPO is a phase-controlled operation, whose frequency can be independent of the cavity delay. Hence, continuous frequency tuning can be achieved without the need for modification of the cavity delay components. 

Here, we constructed an OEPO system in which both the internal cavity phase and the external pump phase can be precisely tuned to stabilize either degenerate or nondegenerate auto-oscillations. Further, by incorporating a magnonic resonator based on a $\rm Y_3Fe_5O_{12}$(YIG) sphere into the OEPO, we demonstrate strong and coherent coupling between YIG's magnon modes and the OEPO's characteristic photon modes. Our system, termed the optoelectronic-magnonic parametric oscillator (OEMPO), bestows additional advantages than using conventional delay-operated OEOs, such as multimode hybridization, high frequency tunability, robust mode stability, and coherent phase operations, which may be used as modular component for constructing future `distributed, hybrid magnonic systems'.        

\section{Experiments}

Figure \ref{fig:scheme} illustrates the experimental setup, which embeds different configurations controlled by the two manual connections at SW$_1$ and SW$_2$. The central build is a standard OEO that is realized by connecting both SW$_1$ and SW$_2$ to their respective position-1s. It includes an electronic sector (consisting of microwave components for amplifying, attenuating, phase tuning, and spectral detection of the signals) and a photonic sector (consisting of E-O/O-E converters and a fiber delay). In particular, for a proof-of-concept, the photonic part can be conveniently implemented by using fiber-optic patch cords and off-the-shelf RF-Over-Fiber (RFoF) systems. 

The dashline-enclosed box on the left is a magnonic resonator based on YIG spin-wave resonances (along with the probe subloop), which can be incorporated into the main OEO loop by switching SW$_2$ to its position-2. The YIG sphere has a nominal diameter of 1.0 mm. By using a pair of circulators, the magnon spectrum can be probed in a counter-flowing subloop without disturbing the main circulation of the signal. The subloop measures the magnon resonances by detecting the rf transmission at a given frequency using a microwave diode and a lock-in amplifier, under a field-modulated ferromagnetic resonance (FMR) scheme \cite{inman2022hybrid,qu2025pump}. Such an augmented system, termed as the opto-electronic-magnonic oscillator (OEMO), has been addressed in an earlier report \cite{xiong2024magnon}.    

The dashline-enclosed box on the right is the electrical parametric pumping module that can be incorporated into the loop by connecting SW$_1$ to its position-2, transforming the system into an OEPO \cite{hao2020optoelectronic}. The key component to the parametric-pumping functionality is the 3-port frequency mixer -- a nonlinear electrical device generating new frequencies due to the second-order nonlinearity. The mixer plays the same role as the optical nonlinear crystal in a standard photonic OPO. When an LO pump signal, $\rm f_p$ is applied, a pair of frequencies, $\rm f_{p1}$ and $\rm f_{p2}$, can be induced and converted into each other inside the mixer, which subsequently enter the OEO cavity loop and form sustained auto-oscillations \cite{hao2020optoelectronic}. Different from the conventional parametric oscillator where the pump provides the gain, the signal gain in this scheme is provided by the loop, which consists of amplifying and phase-tuning rf elements, see Fig.\ref{fig:scheme}. It is noted that such frequency conversion is unconditional (requires no phase-matching as in conventional OPOs), thus stable multi-mode oscillations can be expected.    

Combining both dashline-enclosed boxes (connecting both SW$_1$ and SW$_2$ to their position-2s) then allows the magnon resonator to couple to the parametrically-pumped OEO modes, termed as an opto-electronic-magnonic-parametric oscillator (OEMPO), whose unique spectral characteristics will be addressed below.

\section{Results and Discussions}

To study the OEMPO characteristics, we added two phase shifters, PhS$_1$ and PhS$_2$. The PhS$_1$ (in the loop) tunes the cavity phase and the PhS$_2$ (outside the loop, after the pump) modulates the input pump phase. Both are controlled digitally via external voltage bias, $V_{\varphi1}$ and $V_{\varphi2}$, respectively. 

\begin{figure}[htb]
 \centering
 \includegraphics[width=3.5 in]{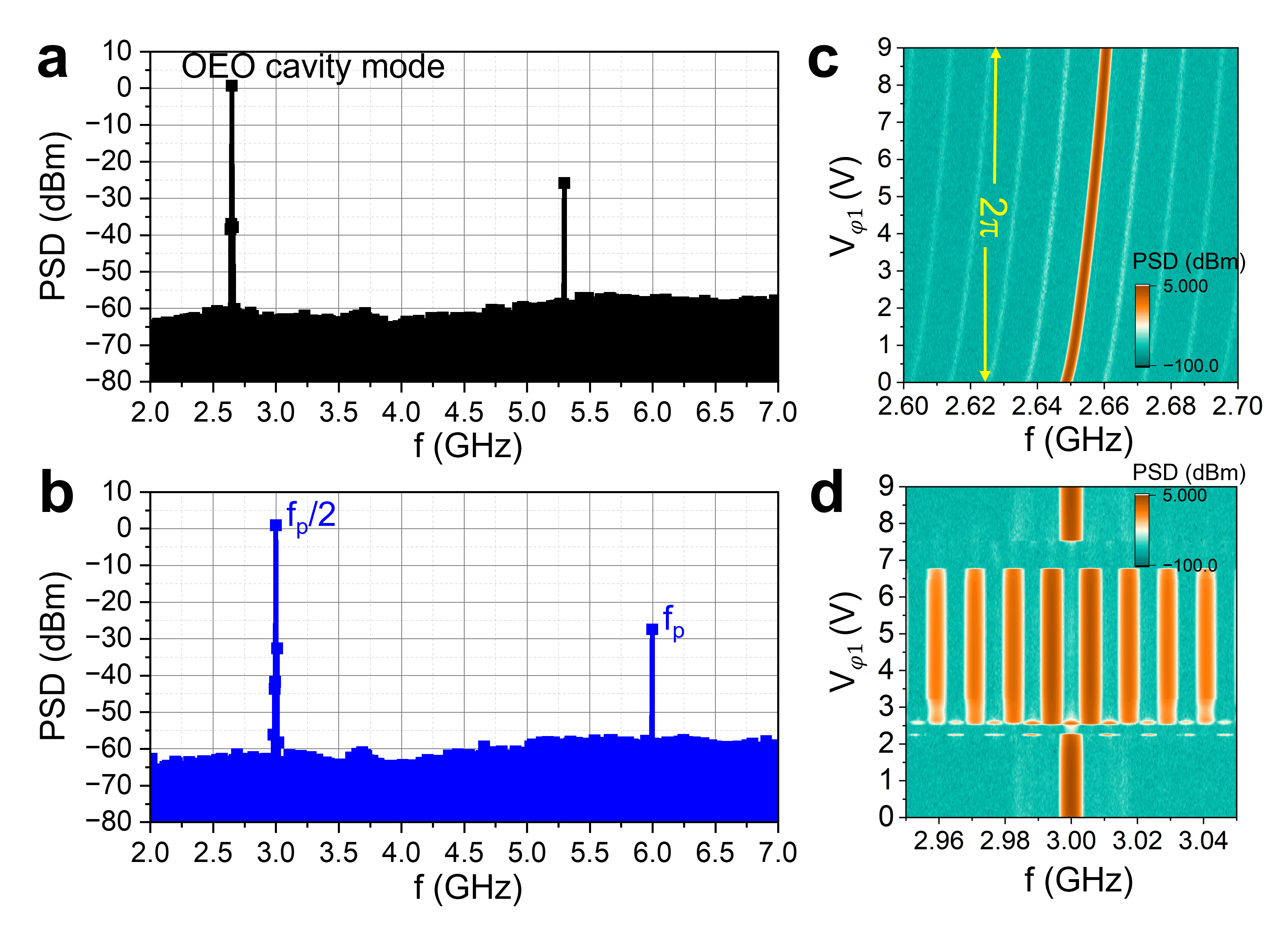}
 \caption{(a) OEO power spectrum showing the intrinsic cavity loop mode, $\rm f_o$, at 2.65 GHz, and the 2nd harmonic mode, 2$\rm f_o$, at 5.3 GHz. (b) OEPO power spectrum showing the half-pump frequency harmonics, $\frac{\rm f_p}{2}$, at 3 GHz, and the pump frequency, $\rm f_p$, at 6 GHz. (c,d) Evolution of the power spectrum by tuning the cavity loop phase, via changing the $V_{\varphi1}$ in PhS$_1$, for the (c) OEO loop mode at $\rm f_o =$2.65 GHz, and (d) OEPO oscillation modes near the half-pump harmonics, $\frac{\rm f_p}{2} = 3$ GHz. } 
 \label{fig:rough}
\end{figure}

Figure \ref{fig:rough} compares the spectral characteristics of the OEPO configuration (with pump) to the standard OEO configuration (without pump), under the setup depicted in Fig. \ref{fig:scheme}. As an example, we elect to use a pump frequency, $\rm f_p$, of 6-GHz, along with the appropriated corresponding bandpass filters (BPF) centered at the half-pump frequency, $\frac{\rm f_p}{2} = 3$ GHz, in the cavity loop. As shown in Fig. \ref{fig:rough}(a), without the pump element (SW$_1$ to position-1), the system regresses to a conventional OEO, with a fundamental eigenmode, $\rm f_o$, appears at 2.65 GHz, stabilized by the cavity's loop delay when the Barkhausen stability criteria is satisfied:
\begin{equation}
    \beta A_\textrm{loop} = 1, \angle \beta A_\textrm{loop} = 2\pi n, n \in \{0,1,2,...\}.
\label{Eq:oeo} 
\end{equation}
\noindent in which $A_\textrm{loop}$ is the total amplification in the loop, $\beta$ is the total loss, and $\angle \beta A_\textrm{loop}$ is the phase accumulated along the loop. Besides the fundamental eigenmode, the 2nd harmonic mode, $2\rm f_o$, is also seen in our measured spectral range, at 5.3 GHz.  

Next, adding the pump element (SW$_1$ to position-2) substantially altered the spectral behavior of the system, see Fig. \ref{fig:rough}(b). The intrinsic cavity eigenmode was significantly suppressed; instead, the system is dominated by the parametric pumping effect: a prominent mode at half of the pump frequency, $\rm \frac{f_p}{2}=3$ GHz, resides inside the loop, besides the pump itself, $\rm f_p$ at 6 GHz, whose amplitude is $\sim 30$ dB lower than for the $\rm \frac{f_p}{2}$ mode. It is noted that for such a parametric pumping process, the power of the pump mode is not a critical parameter, as it only seeds a frequency to the system; the $\rm \frac{f_p}{2}$ mode is the actual circulating mode inside the cavity, and is amplified by the series of in-loop amplifiers. 

Due to the nonlinear pumping mechanism, the OEPO's spectral response to the cavity loop phase (PhS$_1$) is in stark contrast to that of a standard OEO. According to the Barkhausen stability criteria, modulating the loop phase will cause the phase condition, i.e. Eq.\eqref{Eq:oeo} to break down for a prior stabilized OEO eigenmode, and hence, a shift of the eigenmode to a new frequency. Such a behavior was confirmed in our OEO cavity shown in Fig.\ref{fig:rough}(c): the eigenmode frequency gradually changes from 2.65 GHz (at $V_{\varphi1} = 0$V) to 2.66 GHz (at $V_{\varphi1} = 9$V), i.e., approximately a full FSR ($\sim 12$ MHz) across a net phase shift of $\sim 2\pi$.  

On the other hand, the mixer, as the nonlinear medium, produces new signals at the difference of the original frequencies. Therefore, given a LO frequency (pump, $\rm f_p$), a pair of oscillations, $\rm f_{p1}$ and $\rm f_{p2}$, at the RF and IF ports, respectively, will convert into each other back and forth in the electrical mixer, see Fig. \ref{fig:scheme}. Thus, the boundary condition of the OEPO is written as \cite{hao2020optoelectronic}: 
\begin{equation}
    -\rm 2\pi f_{p1,2}\tau + \rm \varphi_{s1,2} + 2\pi N_{1,2} = \varphi_{s2,1},
\label{Eq:oepo_bc}
\end{equation}
\noindent in which $\tau$ is the cavity delay, $\rm \varphi_{s1,2}$ are the initial phases, and $\rm N_{1,2}$ are integers. As a result, the OEPO allows phase jumps, $\rm \varphi_{s1,2}$, as long as they appear in the pair-form. Such phase jump exempts the system from the stringent Barkhausen phase condition, Eq.\eqref{Eq:oeo}, in stabilizing an auto-oscillation mode. Assuming that the additive phase noise in the optoelectronic cavity is negligible, according to Eq.\eqref{Eq:oepo_bc}, the nondegenerate mode pairs are obtained as \cite{hao2020optoelectronic}:
\begin{equation}
    \rm f_{p1}, \rm f_{p2} = \frac{\rm f_{p}}{2} \pm \frac{N \times FSR}{2},  \rm \varphi_{s1}+\rm \varphi_{s2} = -\pi f_{p}\tau+M\pi .
\label{Eq:mode_pair}
\end{equation}
\noindent in which $\rm N = N_1 - N_2$, $\rm M = N_1+N_2$, and $\rm f_{p1} + \rm f_{p2} = \rm f_{p}$. One consequence of the phase jump is that continuous frequency tuning can be achieved, while modification of the cavity delay is not required. In other words, the oscillation is phase-controlled rather than delay-controlled. The mode cluster is determined not only by the cavity delay but also by the LO in the mixer. Modification of either of them would change the mode cluster. Moreover, the oscillating modes must appear symmetrically about half of the LO frequency. The frequency spacing between two cavity modes can be less than the FSR, more precisely, minimally half of the FSR \cite{hao2020optoelectronic}. 

In the case of degenerate oscillation, the frequency converted signal has the same frequency as the input signal, i.e., $\rm f_{p1} = \rm f_{p2}$. The phase-conjugate operation in the parametric process forces the signal to oscillate at half the LO frequency, $\rm \frac{f_{p}}{2}$, regardless of the cavity delay. Therefore, Eq.\eqref{Eq:mode_pair} can be re-written as: 
\begin{equation}
    \rm f_{p1} = \rm f_{p2} = \rm \frac{f_{p}}{2},  \rm \varphi_s = -\frac{\pi \rm f_p\tau}{2}+\frac{\rm M \pi}{2}.
\label{Eq:degenerate} 
\end{equation}
\noindent Different from the conventional oscillator, where the signal has a random phase because oscillation starts from noise, the degenerate oscillation in the OEPO has a specific initial phase, $\rm \varphi_s$, relative to the LO, i.e., the sum phase of the signals before and after the frequency conversion equals the phase of the LO.

\begin{figure}[htb]
 \centering
 \includegraphics[width=3.45 in]{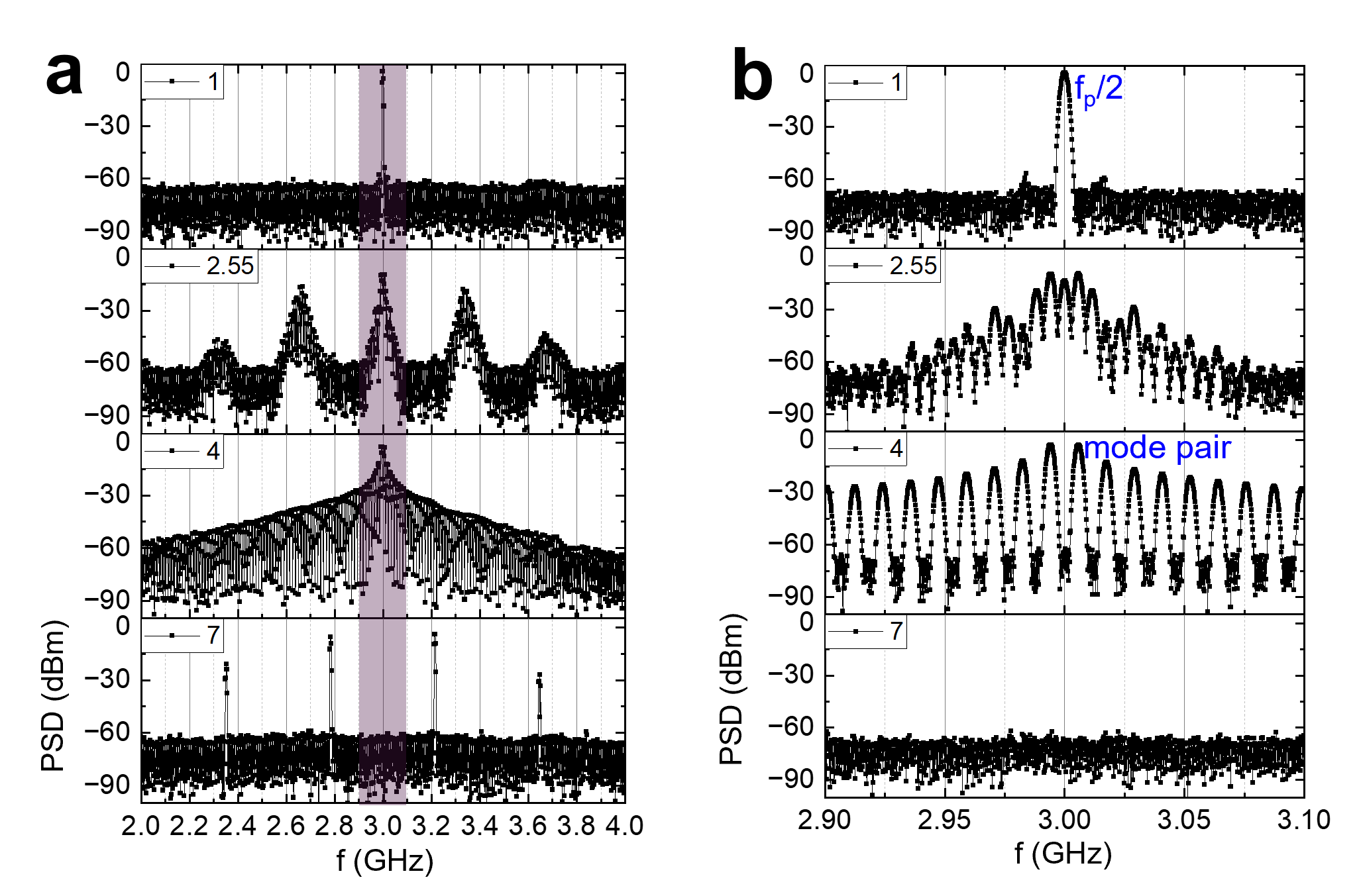}
 \caption{(a) OEPO power spectrum scanned in a broader frequency range from 2 to 4 GHz (2000 MHz span), showing the spectral envelope, at selective, representative $V_{\varphi1}$ values, (top to bottom) 1, 2.55, 4, and 7V. (b) The corresponding fine scan of the power spectrum of the shaded area in (a), from 2.9 to 3.1 GHz (200 MHz span), at the same $V_{\varphi1}$ values. The mode spacing of the mode-pair displayed in (b) at $V_{\varphi1}=4$V is 6 MHz, half of the full FSR (12 MHz). }
 \label{fig:fine}
\end{figure}

The nondegenerate mode-pairs and the degenerate half-frequency mode can be realized by fine tuning the loop phase, PhS$_1$, as illustrated in Fig.\ref{fig:rough}(d). The transition from single mode to mode-pairs occurs around $V_{\varphi1} = 2.5$V, and the single mode appears again at $V_{\varphi1} > 7.5$V. Other than the transition regimes, the mode frequency is quite robust against phase-tuning, owing to the additional phase jump in the parametric process, as in Eq.\eqref{Eq:degenerate}.

To gain more insight into the microscopic features of the phase-tuned OEPO spectrum, we performed fine spectral scans near the $\rm \frac{f_p}{2}$ mode at different $V_{\varphi1}$ values, as shown in Fig. \ref{fig:fine}. In our setup, the delay length is estimated as $L=c\tau = c/$FSR = 16.7 m. The speed of light in the fibers and the microwave cables is $c \sim 2.0 \times10^8$ m/s, so the cavity FSR in our system is determined to be $\sim 12$ MHz. 

At $V_{\varphi1} = 1$V, the degenerate, single-mode oscillation at $\rm \frac{f_p}{2}$ is observed. In this sense, the function and structure of the OEPO are similar to that of an electrical frequency divider, which operates by synchronizing the oscillation frequency of the parametric oscillator to one of the subharmonics of the pump signal. 

\begin{figure}[htb]
 \centering
 \includegraphics[width=3.3 in]{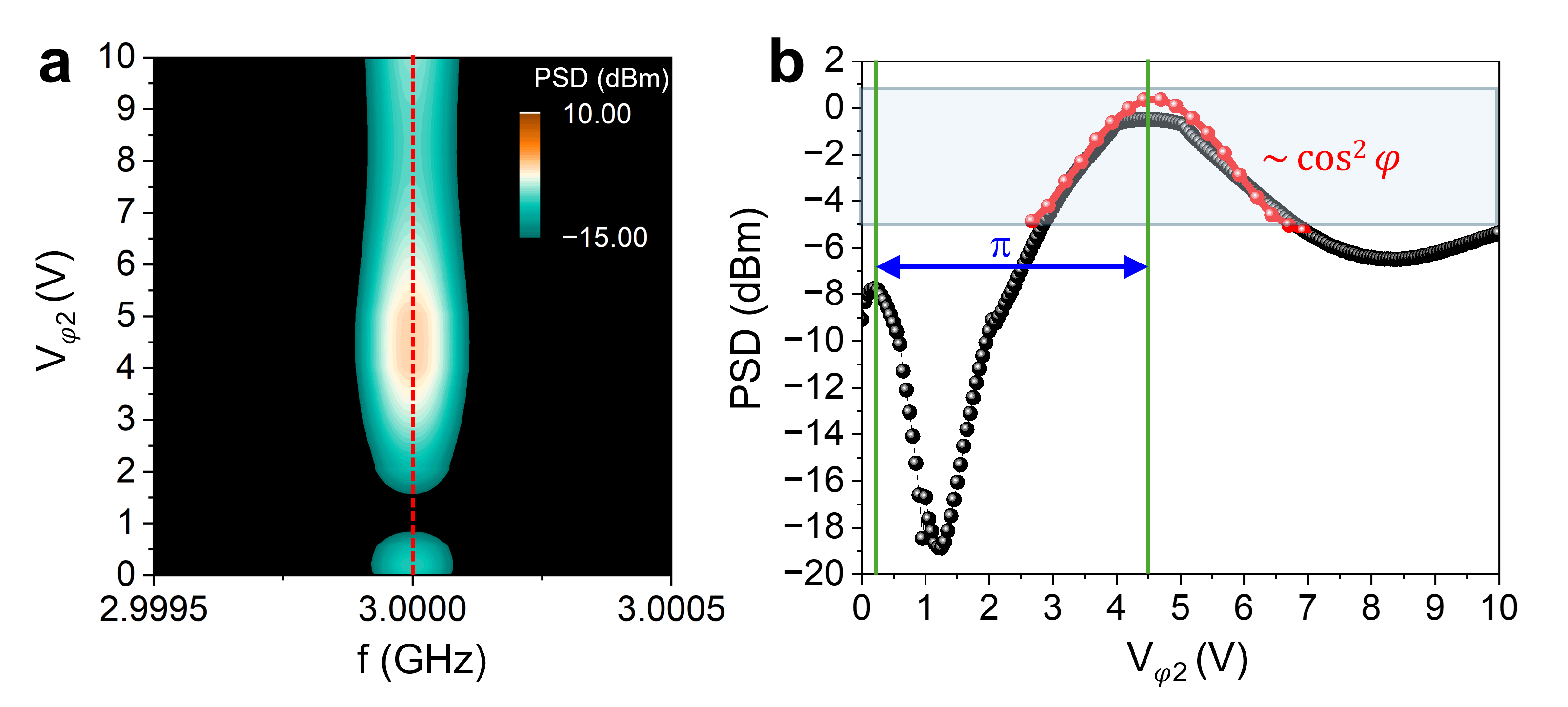}
 \caption{(a) The OEPO power spectrum scanned near the $\rm \frac{f_p}{2}$ mode at different pump phases, via tuning the $V_{\varphi2}$ in the PhS$_2$. The $V_{\varphi1}$ is fixed at 1.89V to stabilize the $\rm \frac{f_p}{2}$ mode during the scan. (b) A line-cut scan at 3 GHz of the power spectrum in (a). A global maximum is observed around $V_{\varphi2} \sim 4.5$V, and a local maximum is observed around $V_{\varphi2} \sim 0.3$V. The phase difference between the local maximum and the global maximum is calculated to be $\sim \pi$. }
 \label{fig:v2}
\end{figure}

\begin{figure*}[htb]
 \centering
 \includegraphics[width=5.8 in]{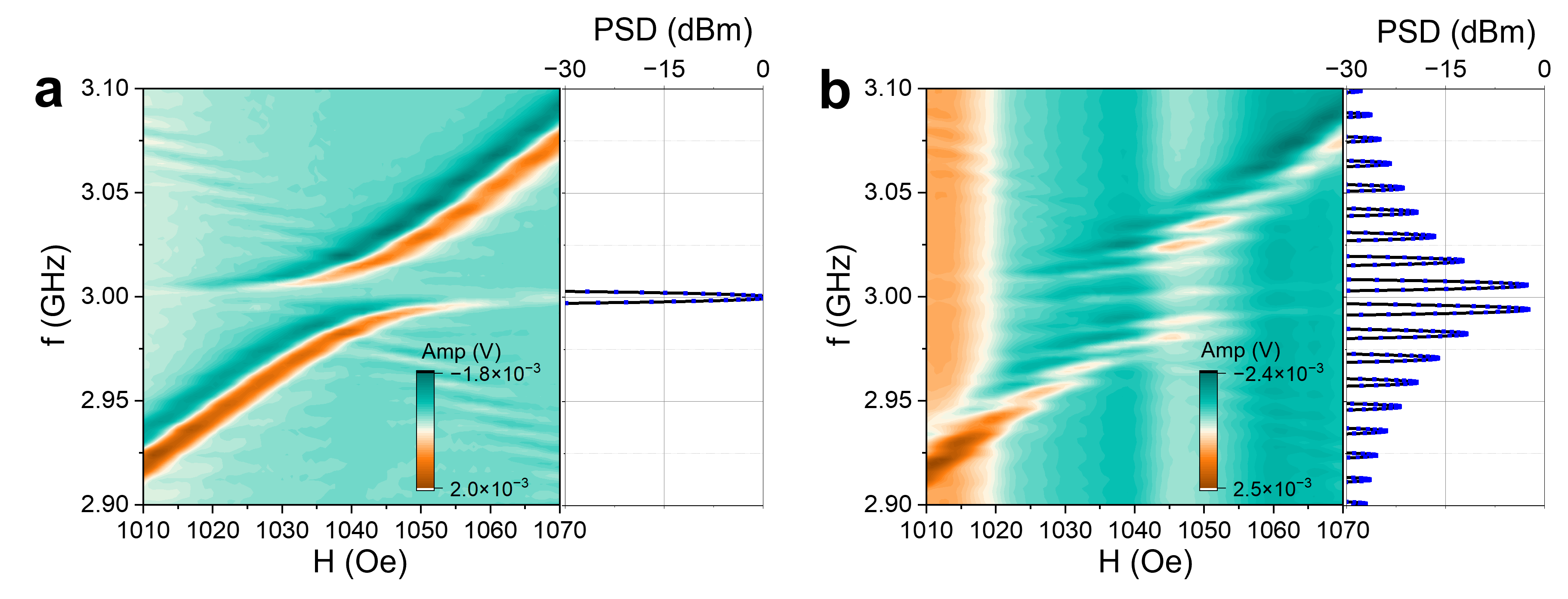}
 \caption{Photon-magnon coupling between the YIG's magnetostatic mode and the OEPO's characteristic photon modes measured via the probe subloop. (a,b) The $f-H$ dispersion spectra of the YIG magnon couples to: (a) the degenerate, single-mode $\rm \frac{f_p}{2}$ auto-oscillation, and (b) the nondegenerate mode-pairs, akin to a photonic frequency comb. The corresponding OEPO power spectrum is also displayed for each case, which is detected using an auxiliary spectrum analyzer. \textcolor{black}{The OEPO spectral response exhibits a long-term stability during the concurrent $f-H$ dispersion measurements. } }
 \label{fig:magn}
\end{figure*}

At $V_{\varphi1} = 2.55$V, the degenerate, single-mode oscillation starts to evolve into mode-pairs, and the harmonics of both the single-mode and mode-pairs co-exist. Further increasing the $V_{\varphi1}$ then stabilizes the mode-pairs, for example, shown in the spectrum at $V_{\varphi1} = 4$V. As the theory predicts, the multimode oscillations are symmetrical about half of the pump frequency, and the minimum mode spacing is half of the cavity FSR, $\sim 6$ MHz. Because the parametric process only locks the sum phase of each mode pair, the frequency and the initial phase of the mode pairs are not unique, hence a frequency comb structure is often result, with the mode spacing being integer numbers of the half-FSR, see Fig.\ref{fig:fine}; however, the two frequency components of each mode pair always have the same power. In fact, the oscillating mode spacing can be much larger than the cavity FSR as long as the initial phase and frequency of the oscillation modes satisfy Eq.\eqref{Eq:mode_pair}. For example, at $V_{\varphi1} = 7$V, the closest stable mode pairs occurs at 2.79 and 3.21 GHz, with a mode spacing of 420 MHz, corresponding to $70 \times \frac{\rm FSR}{2}$. Nevertheless, the oscillation modes of the OEPO must oscillate in pairs and start at the same time with the same amplitude, despite their different spectral envelopes, which is consistent with the theory.  

As discussed earlier, the phase conjugate operation locks the sum phase of the signals before and after the frequency conversion to the injected LO phase. It is therefore worthwhile to examine the effect of the injected LO phase, via the PhS$_2$ installed before the mixer in the pump element. We first tuned the system to the degenerate, single-mode oscillation of $\rm \frac{f_p}{2}=3$ GHz with a $V_{\varphi1}$ value set to 1.89V, and then scanned the power spectrum near $\rm \frac{f_p}{2}$ but at different $V_{\varphi2}$ values in the PhS$_2$. A nontrivial power variation of the $\rm \frac{f_p}{2}$ mode spectrum was observed, upon changing different input phases (PhS$_2$) of the pump $\rm f_p$ (at 6 GHz), as shown in Fig. \ref{fig:v2}(a). An oscillation intensity hotspot was found at $V_{\varphi2} \sim 4.5$V. A line-cut scan at 3 GHz of the power spectrum in Fig. \ref{fig:v2}(a) is plotted in Fig. \ref{fig:v2}(b). Apart from the global maximum at 4.5V, another local maximum was observed around $V_{\varphi2} \sim 0.3$V, and a minimum oscillation is observed at $V_{\varphi2} \sim 1.3$V. The phase difference between the local maximum and the global maximum is calculated to be $\sim \pi$. A curve fit near the global maximum $\sim$4.5V agrees reasonably well with a $\rm cos^2(\varphi)$ dependence. Such a spectral characteristic of the input phase PhS$_2$ further corroborates the parametric process occurring at the nonlinear electrical medium (mixer).

Next, the photon-magnon coupling between the OEPO and an additional magnonic resonator can be studied by exploiting the probe subloop, in which the YIG's magnetostatic (MS) magnon modes are detected by the transmission of a fixed-frequency, low-power rf tone while the bias magnetic field is scanned (using the field-modulated FMR scheme \cite{inman2022hybrid,qu2025pump}), as schematically illustrated in Fig. \ref{fig:scheme}. During this probe measurement, the strong OEPO oscillation signals are also present inside the YIG resonator, acting as a strong pump to the magnon modes. Such a strong pump leads to a peculiar PIM \cite{rao2023unveiling,qu2025pump,rao2025time}, which then causes a pronounced back-action to the probed magnon resonance, depending on the relative detuning of the probe frequency with respect to the OEPO oscillation frequency. 

Figure \ref{fig:magn}(a) depicts the magnon coupling to the OEPO's degenerate, single-mode oscillation at $\rm \frac{f_p}{2}$. The magnon signal is measured as a function of bias magnetic field and probe signal frequency, in the presence of the OEPO's degenerate oscillation at 3 GHz. Notably, the pronounced back-action to the probe signal caused by the OEPO cavity mode leads to a reduction of the amplitude of the YIG's MS magnon mode, which, in turn, suppresses the OEPO pumping process. The net result is then \textcolor{black}{a hybridization between the OEPO's cavity mode and the YIG's MS magnon mode. In the case of strong coupling, the photon-magnon coupling strength $g_{am}$ is greater than the dissipation rate of both photon and magnon subsystems, i.e., $g_{am}>\kappa, g_{am}>\gamma_m$, in which $\kappa$ and $\gamma_m$ denote the damping rates of the OEPO and the MS magnon modes, respectively \cite{awschalom2021quantum,lachance2019hybrid,li2020hybrid}. A coupling cooperativity can be also defined, as $\mathcal{C}=\frac{g_{am}^2}{\kappa\,\gamma_m}$ \cite{bourhill2016ultrahigh}. This leads to} an observation of a spectral anti-crossing feature in the $f - H$ dispersion, centered around the OEPO cavity mode. The anti-crossing gap is estimated to be $\sim 28.8$ MHz in the degenerate case, Fig.\ref{fig:magn}(a). These features can be modeled as the coherent coupling between the PIM and the associated magnon mode through a Jaynes-Cummings type of interaction, a function of the magnon occupancy of the mode, which in turn, depends on the intensity of the cavity mode \cite{rao2023unveiling,qu2025pump}.  

By appropriately tuning the loop phase via the PhS$_1$, we can also hybridize the magnon mode with the OEPO's nondegenerate mode-pairs, as depicted in Fig. \ref{fig:magn}(b). A grand anti-crossing feature is still observed, centered at $\rm \frac{f_p}{2}$, which is superimposed by a series of smaller, individual anti-crossing features due to the hybridization between the magnon mode and the photon frequency comb (series of phase-locked mode-pairs). \textcolor{black}{Due to the dense-packed frequency-comb, it is more challenging to quantify the coupling strength between each comb mode and the magnon mode, however, it is reasonable to believe that the magnon’s coupling to each comb mode is of similar strength, given the common standing-wave nature of the auto-oscillating (loop) modes as well as their similar mode profiles due to their extremely close frequencies. These mode-pairs only differ in their phase boundaries caused by the phase jumps as described in Eq.~\eqref{Eq:mode_pair}, however, prior work has indicated a negligible role of the pump phase to such a Rabi-like process \cite{qu2025pump}, and hence, the resulting magnon-photon coupling strength.}

\section{Analytical Modeling}

To understand the coupled mode spectra, we further developed an analytical model based on coherently coupled oscillators. In our setup, the OEPO resonator acts like a microwave pump that generates an approximately uniform near‑field via the CPW across the YIG sphere, transverse to the dc magnetic field. Its isolated dynamics, i.e.,  degenerate single‑tone and nondegenerate mode‑pairs, are captured by the Hamiltonian:
    
\begin{align}
\frac{H_{\text{a}}}{\hbar}
&= \sum_{m}
\Big( \omega_{+m}\, a_{+m}^\dagger a_{+m} + \omega_{-m}\, a_{-m}^\dagger a_{-m} \Big) \nonumber\\
& + \sum_{m}
\Big( \xi \, e^{+2\pi i \hbar f_p t}\, a_{+m}^\dagger a_{-m}^\dagger
      + \xi^{*}\, e^{-2\pi i \hbar f_p t}\, a_{+m} a_{-m} \Big)\,,
\label{eq:H-one}
\end{align} where 
 the operator $a_{+m}^\dagger$ ($a_{+m}$) creates (annihilates) a photon in the mode above the pump half-frequency by $m\,\frac{\mathrm{FSR}}{2}$, while $a_{-m}^\dagger$ ($a_{-m}$) creates (annihilates) a photon in the symmetric partner mode below the pump half-frequency by the same offset. The complex coefficient $\xi $ parameterizes the mode-pair parametric coupling strength, and the  mode frequencies are given by
\begin{equation}
\omega_{\pm m} = \omega_a \pm m \Delta_a\,,
\end{equation}
with $\omega_a \equiv 2\pi f_p /2$ and $\Delta_a \equiv \pi \mathrm{FSR}$. 
 Since each OEPO mode acts as an independent drive of magnetization dynamics, it is convenient to  replace the pair-creation term in 
Eq.~\eqref{eq:H-one} with an equivalent classical drive on each mode and relabel 
$\omega_{\pm m} \to \omega_{\sigma m}$ for compactness, with $\sigma=\pm$ denoting the 
original sideband index. The Hamiltonian in Eq.~(\ref{eq:H-one}) then becomes:
\begin{align}
\frac{H_{\mathrm{a}}}{\hbar}
=& \sum_{m,\sigma=\pm} \omega_{\sigma m}\, a_{\sigma m}^\dagger a_{\sigma m} \nonumber\\
\quad &+  \sum_{m,\sigma=\pm} 
\left( i\xi e^{-i \hbar \omega_{\sigma m} t} a_{\sigma m}^\dagger + \mathrm{h.c.} \right)\,.
\label{eq:H-one_re}
\end{align}

The interaction between the OEPO modes and the 
$\mathbf{k}=0$ MS magnon mode of YIG $c_0$  can be described by 
\begin{equation}
\frac{H_{ac}}{\hbar} 
= g_{am} \sum_{m,\sigma=\pm} ( a_{\sigma m}^\dagger\, c_{0} )
+ \mathrm{h.c.},
\end{equation}
where $g_{am}$ parameterizes the coupling strength.
When the pump field is sufficiently intense to bring the uniform mode above the Suhl instability threshold \cite{Suhl57}, three–magnon scattering is activated. In this process, the uniform‐mode amplitude $c_{0}$ decays into phase‐correlated magnon pairs $c_{\pm\mathbf{k}}$ at frequency $\omega_0/2$. The three-magnon splitting process and its reverse, i.e., the confluence process are governed by the interaction Hamiltonian \cite{qu2025pump}:
\begin{equation}
H_{3m} = \frac{1}{2}\sum_{\mathbf{k}}
\left(
  \zeta_{\mathbf{k}}\, c_0\, c_{\mathbf{k}}^{*} c_{-\mathbf{k}}^{*}
  + \zeta_{\mathbf{k}}^{*}\, c_0^{*}\, c_{\mathbf{k}} c_{-\mathbf{k}}
\right),
\label{H3m}
\end{equation}
\noindent
where $\zeta_{\mathbf{k}} = \frac{\hbar \gamma M_s}{4} \sin(2\theta)$ is the three--magnon coupling coefficient, 
with $\gamma$ the gyromagnetic ratio, $M_s$ the saturation magnetization, and $\theta$ the angle between the wavevector $\mathbf{k}$ and the applied magnetic field. 
The coupling is maximal at $\theta = 45^{\circ}$, which satisfies the conditions for three--magnon scattering. 
In the case of a spherical sample, $\zeta_{\mathbf{k}}$ is real, so that $\zeta_{\mathbf{k}}^{*} = \zeta_{\mathbf{k}}$.

Introducing the pumped finite-$\bf k$ amplitudes as $c_{\mathbf{k}} = C_{\mathbf{k}} + \delta c_{\mathbf{k}}$ and $c_{-\mathbf{k}} = C_{-\mathbf{k}} + \delta c_{-\mathbf{k}}$, where $C_{\pm\mathbf{k}}$ are the steady (condensate) components  and $\delta c_{\pm\mathbf{k}}$ are small dynamical fluctuations, Eq.~\eqref{H3m} can be rewritten as: 
 \begin{equation}
   \frac{ H_{3m}}{\hbar}=g \big(c_0^{\dagger} p + c_0 p^{\dagger}\big)\,,
\label{eq:H-coupled}
\end{equation}
where 
\begin{align}
p^{\dagger} \equiv \frac{1}{\mathcal{N}} \sum_{\mathbf{k}} \zeta_{\mathbf{k}}
\big(C_{\mathbf{k}}^{\ast}\,\delta c_{-\mathbf{k}}^{\dagger} + C_{-\mathbf{k}}^{\ast}\,\delta c_{\mathbf{k}}^{\dagger}\big)\,,
\end{align}
with
\begin{align}
\mathcal{N}^{2} = \sum_{\mathbf{k}} |\zeta_{\mathbf{k}}|^{2} \big( |C_{\mathbf{k}}|^{2} + |C_{-\mathbf{k}}|^{2} \big)\,,
\label{eq:bright-mode}
\end{align} and $g = \mathcal{N}/(2 \hbar )$. Above threshold, the steady amplitudes scale as $|C_{\mathbf{k}}| \propto \sqrt{\sqrt{P_{\rm in}} - \sqrt{P_{S}}}$ \cite{qu2025pump}, so that: 
\begin{equation}
g^2 \propto \sqrt{P_{\rm in}} - \sqrt{P_{S}} \,,
\label{eq:gbp-scaling}
\end{equation}
where $P_{\rm in}$ is the microwave pump power and $P_{S}$ is the Suhl instability threshold power.

The FMR absorption spectra shown in Fig.~\ref{fig:magn} are proportional to the imaginary part of the dynamical susceptibility of the MS mode. In the absence of any optoelectronic oscillator mode resonant with the MS mode, the latter cannot be driven above the Suhl threshold, and no condensate is formed. In this case, the absorption spectrum exhibits a single peak centered at its characteristic frequency $\omega_0 (H)$.
On the other hand, when an OEPO mode is resonant with the MS mode, the dynamics of small probe–induced fluctuations around the steady pumped state can be conveniently expressed in the frame rotating at frequency $\omega_0$ as
\begin{align}
\dot{a}_{\sigma m} &= -\kappa  a_{\sigma m} 
- i\,g_{am}\, c_{0}, \label{eq:a_sigma_m} \\[4pt]
\dot{c}_{0} &= -\gamma_{m} \, c_{0} 
- i\,g_{am}\, a_{\sigma m} 
- i\,g\, p + f, \label{eq:c0} \\[4pt]
\dot{p} &= - \gamma_{p} \, p 
- i\,g \, c_{0}, \label{eq:p}
\end{align}
where $\kappa$, $\gamma_{m}$ and $\gamma_{p}$ denote, respectively, the damping rates 
of the OEPO, MS magnon and condensate modes, while $f$ is the (weak) probe amplitude. 
\textcolor{black}{Introducing the bright and dark superpositions:
\begin{align}
b_{\sigma m} =\frac{g_{am}\,a_{\sigma m} + g_p\,p}{g_T}, \;  \; 
d_{\sigma m}= \frac{g\,a_{\sigma m} - g_{am}\,p}{g_T},
\end{align}
Eqs.~\eqref{eq:a_sigma_m}--\eqref{eq:p} can be recast as
\begin{align}
\dot b_{\sigma m} &= -\kappa_{\!\mathrm{eff}}\,b_{\sigma m} - \varepsilon\, d_{\sigma m} - i g_T\,c_0, \label{eq:b_dot}\\
\dot d_{\sigma m} &= -\kappa_{\!\mathrm{dark}}\,d_{\sigma m} - \varepsilon\, b_{\sigma m},
 \label{eq:d_dot}\\
\dot c_0 &= -\,\gamma_m\,c_0 - i\,g_T\,b_{\sigma m} + f, \label{eq:c0_dot_bd}
\end{align}
with  \( g_T = \sqrt{g_{am}^2 + g^2} \), $
\kappa_{\!\mathrm{eff}}= (\kappa g_{am}^2 + \gamma_p g^2)/g_T^2$, $
\kappa_{\!\mathrm{dark}}= (\kappa g^2 + \gamma_p g_{am}^2)/g_T^2$, and $
\varepsilon = (\kappa-\gamma_p) g\,g_{am}/g_T^2
$.
Equations~(\ref{eq:b_dot})–(\ref{eq:c0_dot_bd}) show that, while the dark mode \(d_{\sigma m}\) couples dissipatively to  \(b_{\sigma m}\), it remains decoupled from the magnon mode \(c_{0}\), and it is therefore invisible to the FMR probe. In the near-resonant, small-contrast limit \(\lvert\kappa-\gamma_{p}\rvert \ll \kappa+\gamma_{p}\), the bright mode is well approximated as a single lossy channel with effective linewidth \(\kappa_{\mathrm{eff}}\) that hybridizes with \(c_{0}\) at the collective rate $g_T$; the corresponding two-mode cooperativity is then rewritten as
$\mathcal{C}=\frac{g_T^2}{\kappa_{\!\mathrm{eff}}\,\gamma_m}$. 
Fourier transforming Eqs.~\eqref{eq:a_sigma_m}--\eqref{eq:p} and solving for the MS mode response, we find
\begin{align}
   \left[ \gamma_m-i\omega + \frac{g_{am}^2}{\kappa-i\omega} + \frac{g^2}{\gamma_p-i\omega} \right]c_0=f
   \label{eq:d21}
\end{align}
From Eq.~\eqref{eq:d21}, it can be easily shown that, in the limit 
$\kappa, \gamma_{m}, \gamma_{p} \to 0$, the poles of the 
imaginary part of the MS mode susceptibility in the 
laboratory frame reduce to the simple form:
\begin{align}
 \omega = \omega_{0}\pm g_T\,,   
\end{align}
where the splitting  $2 g_T$ corresponds to the anti-crossing gaps shown in Fig.~\ref{fig:magn}.} 
For the degenerate case, an anti-crossing occurs only at the bias field $H$ satisfying $\omega_0 (H)=\omega_a$, in agreement with Fig.~\ref{fig:magn}(a). In contrast, in the nondegenerate case, the resonance condition $\omega_0 (H)=\omega_a \pm m \Delta_a$  can be met for multiple values of $H$,  as  shown in Fig.~\ref{fig:magn}(b).

\section{Conclusion}

In summary, we constructed an OEMPO system incorporating phase-tuning components by which both the internal cavity phase and the external pump phase can be precisely tuned to stabilize either degenerate or nondegenerate auto-oscillations. This way, the photonic part of the OEMPO can be neatly replaced by readily available RFoF systems. Further, by incorporating a magnonic resonator based on a YIG sphere, we demonstrate strong and coherent coupling between YIG's magnon modes and the parametrically pumped, characteristic photon modes. Our system bestows several advantages in the context of integrating hybrid auto-oscillations compared to using conventional delay-operated OEOs:

\textit{First}, unlike standard OEOs that use tunable bandpass filters \cite{pan2024frequency}, phase compensated loops \cite{zhu2020frequency}, or phase-locked loops \cite{zhang2014long} for frequency stabilization, the OEMPO takes an active route towards robust, well-defined frequency injection via a parametric process through an external pump. 

\textit{Second}, due to the phase jumps accommodated at the nonlinear electric medium (mixer), independent controls over the FSR (Barkhausen criteria) and the mode frequency (parametric pump) can be realized, offering unprecedented high tunability to the cavity mode frequencies without changing the cavity delay. 

\textit{Third}, by exploiting a dual-nonlinearity scheme, the OEMPO achieves a magnon-photon mode hybridization through combining the optoelectronic nonlinearity at the mixer (parametric pump) and the magnonic nonlinearity at the YIG (three-magnon scattering), offering a unique testbed for exploring coupled nonlinear phenomena at the hybrid interfaces among electronic, optical, and magnetic platforms \cite{valagiannopoulos2022multistability}. \textcolor{black}{In particular, magnon modes under confined geometries or subject to nonlinear mixing processes can also give rise to a magnonic frequency comb that is controllable by external magnetic field \cite{hula2022spin,wang2024enhancement,christy2025tuning}. This way it is possible to achieve mode hybridization between two interfering frequency combs, that is manifested by an array of nondegenerate, hybridized photon-magnon modes. The coupling at each node is characterized by a unique set of mode-pertaining parameters, including the wavevector \cite{xu2024strong,christy2025tuning}, phase \cite{joseph2024role,gardin2023manifestation}, and mode profile \cite{li2019strong,hou2019strong}. }   

As a result, the system's multimode hybridization, high frequency tunability, robust mode stability, and coherent phase operations may render it useful as a modular component for constructing future distributed, hybrid magnonic systems.

\section*{Acknowledgments} 

The authors acknowledge useful discussions with Yi Li. The experimental work at the UNC-CH was supported by the U.S. Department of Energy, Office of Science, Basic Energy Sciences under Award Number DE-SC0026305. The theoretical work at BC was supported by U.S. National Science Foundation under Grant No. ECCS-2337713.

\section*{AUTHOR DECLARATIONS} 

\subsection*{Conflict of Interest} 

The authors have no conflicts of interest to disclose. 

\subsection*{Author Contributions} 

\textbf{Junming Wu:} Data curation (lead); Formal analysis (supporting); Investigation (supporting); Methodology (supporting); Visualization (supporting); Writing–original draft (supporting); Writing– review $\&$ editing (equal).
\textbf{Shihao Zhou:} Data curation (supporting); Formal analysis (supporting); Methodology (supporting); Visualization (supporting); Writing–original draft (supporting); Writing– review $\&$ editing (equal).
\textbf{Benedetta Flebus:} Conceptualization (equal); Formal analysis (lead); Investigation (supporting); Methodology (supporting); Visualization (supporting); Writing–original draft (equal); Writing– review $\&$ editing (equal).
\textbf{Wei Zhang:} Conceptualization (equal); Data curation (supporting); Formal analysis (supporting); Investigation (lead); Methodology (lead); Visualization (lead); Writing–original draft (equal); Writing– review $\&$ editing (equal).

\section*{DATA AVAILABILITY} 

Data supporting the findings of this study are available from the corresponding authors upon reasonable request. 

\section*{References}  
\bibliography{ref}

\begin{thebibliography}{42}%
\makeatletter
\providecommand \@ifxundefined [1]{%
 \@ifx{#1\undefined}
}%
\providecommand \@ifnum [1]{%
 \ifnum #1\expandafter \@firstoftwo
 \else \expandafter \@secondoftwo
 \fi
}%
\providecommand \@ifx [1]{%
 \ifx #1\expandafter \@firstoftwo
 \else \expandafter \@secondoftwo
 \fi
}%
\providecommand \natexlab [1]{#1}%
\providecommand \enquote  [1]{``#1''}%
\providecommand \bibnamefont  [1]{#1}%
\providecommand \bibfnamefont [1]{#1}%
\providecommand \citenamefont [1]{#1}%
\providecommand \href@noop [0]{\@secondoftwo}%
\providecommand \href [0]{\begingroup \@sanitize@url \@href}%
\providecommand \@href[1]{\@@startlink{#1}\@@href}%
\providecommand \@@href[1]{\endgroup#1\@@endlink}%
\providecommand \@sanitize@url [0]{\catcode `\\12\catcode `\$12\catcode `\&12\catcode `\#12\catcode `\^12\catcode `\_12\catcode `\%12\relax}%
\providecommand \@@startlink[1]{}%
\providecommand \@@endlink[0]{}%
\providecommand \url  [0]{\begingroup\@sanitize@url \@url }%
\providecommand \@url [1]{\endgroup\@href {#1}{\urlprefix }}%
\providecommand \urlprefix  [0]{URL }%
\providecommand \Eprint [0]{\href }%
\providecommand \doibase [0]{http://dx.doi.org/}%
\providecommand \selectlanguage [0]{\@gobble}%
\providecommand \bibinfo  [0]{\@secondoftwo}%
\providecommand \bibfield  [0]{\@secondoftwo}%
\providecommand \translation [1]{[#1]}%
\providecommand \BibitemOpen [0]{}%
\providecommand \bibitemStop [0]{}%
\providecommand \bibitemNoStop [0]{.\EOS\space}%
\providecommand \EOS [0]{\spacefactor3000\relax}%
\providecommand \BibitemShut  [1]{\csname bibitem#1\endcsname}%
\let\auto@bib@innerbib\@empty
\bibitem [{\citenamefont {Awschalom}\ \emph {et~al.}(2021{\natexlab{a}})\citenamefont {Awschalom}, \citenamefont {Du}, \citenamefont {He}, \citenamefont {Heremans}, \citenamefont {Hoffmann}, \citenamefont {Hou}, \citenamefont {Kurebayashi}, \citenamefont {Li}, \citenamefont {Liu}, \citenamefont {Novosad} \emph {et~al.}}]{awschalom2021quantum}%
  \BibitemOpen
  \bibfield  {author} {\bibinfo {author} {\bibfnamefont {D.~D.}\ \bibnamefont {Awschalom}}, \bibinfo {author} {\bibfnamefont {C.~R.}\ \bibnamefont {Du}}, \bibinfo {author} {\bibfnamefont {R.}~\bibnamefont {He}}, \bibinfo {author} {\bibfnamefont {F.~J.}\ \bibnamefont {Heremans}}, \bibinfo {author} {\bibfnamefont {A.}~\bibnamefont {Hoffmann}}, \bibinfo {author} {\bibfnamefont {J.}~\bibnamefont {Hou}}, \bibinfo {author} {\bibfnamefont {H.}~\bibnamefont {Kurebayashi}}, \bibinfo {author} {\bibfnamefont {Y.}~\bibnamefont {Li}}, \bibinfo {author} {\bibfnamefont {L.}~\bibnamefont {Liu}}, \bibinfo {author} {\bibfnamefont {V.}~\bibnamefont {Novosad}},  \emph {et~al.},\ }\bibfield  {title} {\enquote {\bibinfo {title} {Quantum engineering with hybrid magnonic systems and materials},}\ }\href@noop {} {\bibfield  {journal} {\bibinfo  {journal} {IEEE Transactions on Quantum Engineering}\ }\textbf {\bibinfo {volume} {2}},\ \bibinfo {pages} {1--36} (\bibinfo {year} {2021}{\natexlab{a}})}\BibitemShut {NoStop}%
\bibitem [{\citenamefont {Lachance-Quirion}\ \emph {et~al.}(2019)\citenamefont {Lachance-Quirion}, \citenamefont {Tabuchi}, \citenamefont {Gloppe}, \citenamefont {Usami},\ and\ \citenamefont {Nakamura}}]{lachance2019hybrid}%
  \BibitemOpen
  \bibfield  {author} {\bibinfo {author} {\bibfnamefont {D.}~\bibnamefont {Lachance-Quirion}}, \bibinfo {author} {\bibfnamefont {Y.}~\bibnamefont {Tabuchi}}, \bibinfo {author} {\bibfnamefont {A.}~\bibnamefont {Gloppe}}, \bibinfo {author} {\bibfnamefont {K.}~\bibnamefont {Usami}}, \ and\ \bibinfo {author} {\bibfnamefont {Y.}~\bibnamefont {Nakamura}},\ }\bibfield  {title} {\enquote {\bibinfo {title} {Hybrid quantum systems based on magnonics},}\ }\href@noop {} {\bibfield  {journal} {\bibinfo  {journal} {Applied Physics Express}\ }\textbf {\bibinfo {volume} {12}},\ \bibinfo {pages} {070101} (\bibinfo {year} {2019})}\BibitemShut {NoStop}%
\bibitem [{\citenamefont {Li}\ \emph {et~al.}(2020)\citenamefont {Li}, \citenamefont {Zhang}, \citenamefont {Tyberkevych}, \citenamefont {Kwok}, \citenamefont {Hoffmann},\ and\ \citenamefont {Novosad}}]{li2020hybrid}%
  \BibitemOpen
  \bibfield  {author} {\bibinfo {author} {\bibfnamefont {Y.}~\bibnamefont {Li}}, \bibinfo {author} {\bibfnamefont {W.}~\bibnamefont {Zhang}}, \bibinfo {author} {\bibfnamefont {V.}~\bibnamefont {Tyberkevych}}, \bibinfo {author} {\bibfnamefont {W.-K.}\ \bibnamefont {Kwok}}, \bibinfo {author} {\bibfnamefont {A.}~\bibnamefont {Hoffmann}}, \ and\ \bibinfo {author} {\bibfnamefont {V.}~\bibnamefont {Novosad}},\ }\bibfield  {title} {\enquote {\bibinfo {title} {Hybrid magnonics: Physics, circuits, and applications for coherent information processing},}\ }\href@noop {} {\bibfield  {journal} {\bibinfo  {journal} {Journal of Applied Physics}\ }\textbf {\bibinfo {volume} {128}} (\bibinfo {year} {2020})}\BibitemShut {NoStop}%
\bibitem [{\citenamefont {Yuan}\ \emph {et~al.}(2022)\citenamefont {Yuan}, \citenamefont {Cao}, \citenamefont {Kamra}, \citenamefont {Duine},\ and\ \citenamefont {Yan}}]{yuan2022quantum}%
  \BibitemOpen
  \bibfield  {author} {\bibinfo {author} {\bibfnamefont {H.}~\bibnamefont {Yuan}}, \bibinfo {author} {\bibfnamefont {Y.}~\bibnamefont {Cao}}, \bibinfo {author} {\bibfnamefont {A.}~\bibnamefont {Kamra}}, \bibinfo {author} {\bibfnamefont {R.~A.}\ \bibnamefont {Duine}}, \ and\ \bibinfo {author} {\bibfnamefont {P.}~\bibnamefont {Yan}},\ }\bibfield  {title} {\enquote {\bibinfo {title} {Quantum magnonics: When magnon spintronics meets quantum information science},}\ }\href@noop {} {\bibfield  {journal} {\bibinfo  {journal} {Physics Reports}\ }\textbf {\bibinfo {volume} {965}},\ \bibinfo {pages} {1--74} (\bibinfo {year} {2022})}\BibitemShut {NoStop}%
\bibitem [{\citenamefont {Flebus}\ \emph {et~al.}(2024)\citenamefont {Flebus}, \citenamefont {Grundler}, \citenamefont {Rana}, \citenamefont {Otani}, \citenamefont {Barsukov}, \citenamefont {Barman}, \citenamefont {Gubbiotti}, \citenamefont {Landeros}, \citenamefont {Akerman}, \citenamefont {Ebels} \emph {et~al.}}]{flebus20242024}%
  \BibitemOpen
  \bibfield  {author} {\bibinfo {author} {\bibfnamefont {B.}~\bibnamefont {Flebus}}, \bibinfo {author} {\bibfnamefont {D.}~\bibnamefont {Grundler}}, \bibinfo {author} {\bibfnamefont {B.}~\bibnamefont {Rana}}, \bibinfo {author} {\bibfnamefont {Y.}~\bibnamefont {Otani}}, \bibinfo {author} {\bibfnamefont {I.}~\bibnamefont {Barsukov}}, \bibinfo {author} {\bibfnamefont {A.}~\bibnamefont {Barman}}, \bibinfo {author} {\bibfnamefont {G.}~\bibnamefont {Gubbiotti}}, \bibinfo {author} {\bibfnamefont {P.}~\bibnamefont {Landeros}}, \bibinfo {author} {\bibfnamefont {J.}~\bibnamefont {Akerman}}, \bibinfo {author} {\bibfnamefont {U.~S.}\ \bibnamefont {Ebels}},  \emph {et~al.},\ }\bibfield  {title} {\enquote {\bibinfo {title} {The 2024 magnonics roadmap},}\ }\href@noop {} {\bibfield  {journal} {\bibinfo  {journal} {Journal of Physics: Condensed Matter}\ } (\bibinfo {year} {2024})}\BibitemShut {NoStop}%
\bibitem [{\citenamefont {Chumak}\ \emph {et~al.}(2022)\citenamefont {Chumak}, \citenamefont {Kabos}, \citenamefont {Wu}, \citenamefont {Abert}, \citenamefont {Adelmann}, \citenamefont {Adeyeye}, \citenamefont {{\AA}kerman}, \citenamefont {Aliev}, \citenamefont {Anane}, \citenamefont {Awad} \emph {et~al.}}]{chumak2022advances}%
  \BibitemOpen
  \bibfield  {author} {\bibinfo {author} {\bibfnamefont {A.~V.}\ \bibnamefont {Chumak}}, \bibinfo {author} {\bibfnamefont {P.}~\bibnamefont {Kabos}}, \bibinfo {author} {\bibfnamefont {M.}~\bibnamefont {Wu}}, \bibinfo {author} {\bibfnamefont {C.}~\bibnamefont {Abert}}, \bibinfo {author} {\bibfnamefont {C.}~\bibnamefont {Adelmann}}, \bibinfo {author} {\bibfnamefont {A.}~\bibnamefont {Adeyeye}}, \bibinfo {author} {\bibfnamefont {J.}~\bibnamefont {{\AA}kerman}}, \bibinfo {author} {\bibfnamefont {F.~G.}\ \bibnamefont {Aliev}}, \bibinfo {author} {\bibfnamefont {A.}~\bibnamefont {Anane}}, \bibinfo {author} {\bibfnamefont {A.}~\bibnamefont {Awad}},  \emph {et~al.},\ }\bibfield  {title} {\enquote {\bibinfo {title} {Advances in magnetics roadmap on spin-wave computing},}\ }\href@noop {} {\bibfield  {journal} {\bibinfo  {journal} {IEEE Transactions on Magnetics}\ }\textbf {\bibinfo {volume} {58}},\ \bibinfo {pages} {1--72} (\bibinfo {year} {2022})}\BibitemShut {NoStop}%
\bibitem [{\citenamefont {Xiong}\ \emph {et~al.}(2025)\citenamefont {Xiong}, \citenamefont {Christy}, \citenamefont {Li}, \citenamefont {Sun}, \citenamefont {Comstock}, \citenamefont {Wu}, \citenamefont {Lopez}, \citenamefont {Lei}, \citenamefont {Sun}, \citenamefont {Cahoon} \emph {et~al.}}]{xiong2025photon}%
  \BibitemOpen
  \bibfield  {author} {\bibinfo {author} {\bibfnamefont {Y.}~\bibnamefont {Xiong}}, \bibinfo {author} {\bibfnamefont {A.}~\bibnamefont {Christy}}, \bibinfo {author} {\bibfnamefont {Y.}~\bibnamefont {Li}}, \bibinfo {author} {\bibfnamefont {R.}~\bibnamefont {Sun}}, \bibinfo {author} {\bibfnamefont {A.~H.}\ \bibnamefont {Comstock}}, \bibinfo {author} {\bibfnamefont {J.}~\bibnamefont {Wu}}, \bibinfo {author} {\bibfnamefont {R.}~\bibnamefont {Lopez}}, \bibinfo {author} {\bibfnamefont {S.}~\bibnamefont {Lei}}, \bibinfo {author} {\bibfnamefont {D.}~\bibnamefont {Sun}}, \bibinfo {author} {\bibfnamefont {J.~F.}\ \bibnamefont {Cahoon}},  \emph {et~al.},\ }\bibfield  {title} {\enquote {\bibinfo {title} {Photon-magnon coupling using gain-assisted spoof-localized surface plasmons},}\ }\href@noop {} {\bibfield  {journal} {\bibinfo  {journal} {Optics Express}\ }\textbf {\bibinfo {volume} {33}},\ \bibinfo {pages} {16809--16819} (\bibinfo {year} {2025})}\BibitemShut {NoStop}%
\bibitem [{\citenamefont {Xiong}\ \emph {et~al.}(2024{\natexlab{a}})\citenamefont {Xiong}, \citenamefont {Christy}, \citenamefont {Yan}, \citenamefont {Pishehvar}, \citenamefont {Mahdi}, \citenamefont {Wu}, \citenamefont {Cahoon}, \citenamefont {Yang}, \citenamefont {Hamilton}, \citenamefont {Zhang} \emph {et~al.}}]{xiong2024hybrid}%
  \BibitemOpen
  \bibfield  {author} {\bibinfo {author} {\bibfnamefont {Y.}~\bibnamefont {Xiong}}, \bibinfo {author} {\bibfnamefont {A.}~\bibnamefont {Christy}}, \bibinfo {author} {\bibfnamefont {Z.}~\bibnamefont {Yan}}, \bibinfo {author} {\bibfnamefont {A.}~\bibnamefont {Pishehvar}}, \bibinfo {author} {\bibfnamefont {M.}~\bibnamefont {Mahdi}}, \bibinfo {author} {\bibfnamefont {J.}~\bibnamefont {Wu}}, \bibinfo {author} {\bibfnamefont {J.~F.}\ \bibnamefont {Cahoon}}, \bibinfo {author} {\bibfnamefont {B.}~\bibnamefont {Yang}}, \bibinfo {author} {\bibfnamefont {M.~C.}\ \bibnamefont {Hamilton}}, \bibinfo {author} {\bibfnamefont {X.}~\bibnamefont {Zhang}},  \emph {et~al.},\ }\bibfield  {title} {\enquote {\bibinfo {title} {Hybrid magnonics with localized spoof surface-plasmon polaritons},}\ }\href@noop {} {\bibfield  {journal} {\bibinfo  {journal} {Physical Review Applied}\ }\textbf {\bibinfo {volume} {22}},\ \bibinfo {pages} {034009} (\bibinfo {year} {2024}{\natexlab{a}})}\BibitemShut {NoStop}%
\bibitem [{\citenamefont {Xiong}\ \emph {et~al.}(2024{\natexlab{b}})\citenamefont {Xiong}, \citenamefont {Christy}, \citenamefont {Dong}, \citenamefont {Comstock}, \citenamefont {Sun}, \citenamefont {Li}, \citenamefont {Cahoon}, \citenamefont {Yang},\ and\ \citenamefont {Zhang}}]{xiong2024combinatorial}%
  \BibitemOpen
  \bibfield  {author} {\bibinfo {author} {\bibfnamefont {Y.}~\bibnamefont {Xiong}}, \bibinfo {author} {\bibfnamefont {A.}~\bibnamefont {Christy}}, \bibinfo {author} {\bibfnamefont {Y.}~\bibnamefont {Dong}}, \bibinfo {author} {\bibfnamefont {A.~H.}\ \bibnamefont {Comstock}}, \bibinfo {author} {\bibfnamefont {D.}~\bibnamefont {Sun}}, \bibinfo {author} {\bibfnamefont {Y.}~\bibnamefont {Li}}, \bibinfo {author} {\bibfnamefont {J.~F.}\ \bibnamefont {Cahoon}}, \bibinfo {author} {\bibfnamefont {B.}~\bibnamefont {Yang}}, \ and\ \bibinfo {author} {\bibfnamefont {W.}~\bibnamefont {Zhang}},\ }\bibfield  {title} {\enquote {\bibinfo {title} {Combinatorial split-ring and spiral metaresonator for efficient magnon-photon coupling},}\ }\href@noop {} {\bibfield  {journal} {\bibinfo  {journal} {Phys. Rev. Appl.}\ }\textbf {\bibinfo {volume} {21}},\ \bibinfo {pages} {034034} (\bibinfo {year} {2024}{\natexlab{b}})}\BibitemShut {NoStop}%
\bibitem [{\citenamefont {Xu}\ \emph {et~al.}(2024{\natexlab{a}})\citenamefont {Xu}, \citenamefont {Zhong}, \citenamefont {Zhuang}, \citenamefont {Qian}, \citenamefont {Jiang}, \citenamefont {Pishehvar}, \citenamefont {Han}, \citenamefont {Jin}, \citenamefont {Jornet}, \citenamefont {Zhen} \emph {et~al.}}]{xu2024slow}%
  \BibitemOpen
  \bibfield  {author} {\bibinfo {author} {\bibfnamefont {J.}~\bibnamefont {Xu}}, \bibinfo {author} {\bibfnamefont {C.}~\bibnamefont {Zhong}}, \bibinfo {author} {\bibfnamefont {S.}~\bibnamefont {Zhuang}}, \bibinfo {author} {\bibfnamefont {C.}~\bibnamefont {Qian}}, \bibinfo {author} {\bibfnamefont {Y.}~\bibnamefont {Jiang}}, \bibinfo {author} {\bibfnamefont {A.}~\bibnamefont {Pishehvar}}, \bibinfo {author} {\bibfnamefont {X.}~\bibnamefont {Han}}, \bibinfo {author} {\bibfnamefont {D.}~\bibnamefont {Jin}}, \bibinfo {author} {\bibfnamefont {J.~M.}\ \bibnamefont {Jornet}}, \bibinfo {author} {\bibfnamefont {B.}~\bibnamefont {Zhen}},  \emph {et~al.},\ }\bibfield  {title} {\enquote {\bibinfo {title} {Slow-wave hybrid magnonics},}\ }\href@noop {} {\bibfield  {journal} {\bibinfo  {journal} {Physical Review Letters}\ }\textbf {\bibinfo {volume} {132}},\ \bibinfo {pages} {116701} (\bibinfo {year} {2024}{\natexlab{a}})}\BibitemShut {NoStop}%
\bibitem [{\citenamefont {Li}\ \emph {et~al.}(2021)\citenamefont {Li}, \citenamefont {Zhao}, \citenamefont {Zhang}, \citenamefont {Hoffmann},\ and\ \citenamefont {Novosad}}]{li2021advances}%
  \BibitemOpen
  \bibfield  {author} {\bibinfo {author} {\bibfnamefont {Y.}~\bibnamefont {Li}}, \bibinfo {author} {\bibfnamefont {C.}~\bibnamefont {Zhao}}, \bibinfo {author} {\bibfnamefont {W.}~\bibnamefont {Zhang}}, \bibinfo {author} {\bibfnamefont {A.}~\bibnamefont {Hoffmann}}, \ and\ \bibinfo {author} {\bibfnamefont {V.}~\bibnamefont {Novosad}},\ }\bibfield  {title} {\enquote {\bibinfo {title} {Advances in coherent coupling between magnons and acoustic phonons},}\ }\href@noop {} {\bibfield  {journal} {\bibinfo  {journal} {APL Materials}\ }\textbf {\bibinfo {volume} {9}} (\bibinfo {year} {2021})}\BibitemShut {NoStop}%
\bibitem [{\citenamefont {Wang}\ \emph {et~al.}(2024{\natexlab{a}})\citenamefont {Wang}, \citenamefont {Csaba}, \citenamefont {Verba}, \citenamefont {Chumak},\ and\ \citenamefont {Pirro}}]{wang2024nanoscale}%
  \BibitemOpen
  \bibfield  {author} {\bibinfo {author} {\bibfnamefont {Q.}~\bibnamefont {Wang}}, \bibinfo {author} {\bibfnamefont {G.}~\bibnamefont {Csaba}}, \bibinfo {author} {\bibfnamefont {R.}~\bibnamefont {Verba}}, \bibinfo {author} {\bibfnamefont {A.~V.}\ \bibnamefont {Chumak}}, \ and\ \bibinfo {author} {\bibfnamefont {P.}~\bibnamefont {Pirro}},\ }\bibfield  {title} {\enquote {\bibinfo {title} {Nanoscale magnonic networks},}\ }\href@noop {} {\bibfield  {journal} {\bibinfo  {journal} {Physical Review Applied}\ }\textbf {\bibinfo {volume} {21}},\ \bibinfo {pages} {040503} (\bibinfo {year} {2024}{\natexlab{a}})}\BibitemShut {NoStop}%
\bibitem [{\citenamefont {Awschalom}\ \emph {et~al.}(2021{\natexlab{b}})\citenamefont {Awschalom}, \citenamefont {Berggren}, \citenamefont {Bernien}, \citenamefont {Bhave}, \citenamefont {Carr}, \citenamefont {Davids}, \citenamefont {Economou}, \citenamefont {Englund}, \citenamefont {Faraon}, \citenamefont {Fejer} \emph {et~al.}}]{awschalom2021development}%
  \BibitemOpen
  \bibfield  {author} {\bibinfo {author} {\bibfnamefont {D.}~\bibnamefont {Awschalom}}, \bibinfo {author} {\bibfnamefont {K.~K.}\ \bibnamefont {Berggren}}, \bibinfo {author} {\bibfnamefont {H.}~\bibnamefont {Bernien}}, \bibinfo {author} {\bibfnamefont {S.}~\bibnamefont {Bhave}}, \bibinfo {author} {\bibfnamefont {L.~D.}\ \bibnamefont {Carr}}, \bibinfo {author} {\bibfnamefont {P.}~\bibnamefont {Davids}}, \bibinfo {author} {\bibfnamefont {S.~E.}\ \bibnamefont {Economou}}, \bibinfo {author} {\bibfnamefont {D.}~\bibnamefont {Englund}}, \bibinfo {author} {\bibfnamefont {A.}~\bibnamefont {Faraon}}, \bibinfo {author} {\bibfnamefont {M.}~\bibnamefont {Fejer}},  \emph {et~al.},\ }\bibfield  {title} {\enquote {\bibinfo {title} {Development of quantum interconnects (quics) for next-generation information technologies},}\ }\href@noop {} {\bibfield  {journal} {\bibinfo  {journal} {Prx Quantum}\ }\textbf {\bibinfo {volume} {2}},\ \bibinfo {pages} {017002} (\bibinfo {year} {2021}{\natexlab{b}})}\BibitemShut {NoStop}%
\bibitem [{\citenamefont {Rao}\ \emph {et~al.}(2023{\natexlab{a}})\citenamefont {Rao}, \citenamefont {Wang}, \citenamefont {Yao}, \citenamefont {Chen}, \citenamefont {Zhao},\ and\ \citenamefont {Lu}}]{rao2023meterscale}%
  \BibitemOpen
  \bibfield  {author} {\bibinfo {author} {\bibfnamefont {J.}~\bibnamefont {Rao}}, \bibinfo {author} {\bibfnamefont {C.}~\bibnamefont {Wang}}, \bibinfo {author} {\bibfnamefont {B.}~\bibnamefont {Yao}}, \bibinfo {author} {\bibfnamefont {Z.}~\bibnamefont {Chen}}, \bibinfo {author} {\bibfnamefont {K.}~\bibnamefont {Zhao}}, \ and\ \bibinfo {author} {\bibfnamefont {W.}~\bibnamefont {Lu}},\ }\bibfield  {title} {\enquote {\bibinfo {title} {Meterscale strong coupling between magnons and photons},}\ }\href@noop {} {\bibfield  {journal} {\bibinfo  {journal} {Physical Review Letters}\ }\textbf {\bibinfo {volume} {131}},\ \bibinfo {pages} {106702} (\bibinfo {year} {2023}{\natexlab{a}})}\BibitemShut {NoStop}%
\bibitem [{\citenamefont {Yang}\ \emph {et~al.}(2024)\citenamefont {Yang}, \citenamefont {Yao}, \citenamefont {Xiao}, \citenamefont {Fong}, \citenamefont {Lau},\ and\ \citenamefont {Hu}}]{yang2024anomalous}%
  \BibitemOpen
  \bibfield  {author} {\bibinfo {author} {\bibfnamefont {Y.}~\bibnamefont {Yang}}, \bibinfo {author} {\bibfnamefont {J.}~\bibnamefont {Yao}}, \bibinfo {author} {\bibfnamefont {Y.}~\bibnamefont {Xiao}}, \bibinfo {author} {\bibfnamefont {P.-T.}\ \bibnamefont {Fong}}, \bibinfo {author} {\bibfnamefont {H.-K.}\ \bibnamefont {Lau}}, \ and\ \bibinfo {author} {\bibfnamefont {C.-M.}\ \bibnamefont {Hu}},\ }\bibfield  {title} {\enquote {\bibinfo {title} {Anomalous long-distance coherence in critically driven cavity magnonics},}\ }\href@noop {} {\bibfield  {journal} {\bibinfo  {journal} {Physical Review Letters}\ }\textbf {\bibinfo {volume} {132}},\ \bibinfo {pages} {206902} (\bibinfo {year} {2024})}\BibitemShut {NoStop}%
\bibitem [{\citenamefont {Yao}\ \emph {et~al.}(2023)\citenamefont {Yao}, \citenamefont {Gui}, \citenamefont {Rao}, \citenamefont {Zhang}, \citenamefont {Lu},\ and\ \citenamefont {Hu}}]{yao2023coherent}%
  \BibitemOpen
  \bibfield  {author} {\bibinfo {author} {\bibfnamefont {B.}~\bibnamefont {Yao}}, \bibinfo {author} {\bibfnamefont {Y.}~\bibnamefont {Gui}}, \bibinfo {author} {\bibfnamefont {J.}~\bibnamefont {Rao}}, \bibinfo {author} {\bibfnamefont {Y.}~\bibnamefont {Zhang}}, \bibinfo {author} {\bibfnamefont {W.}~\bibnamefont {Lu}}, \ and\ \bibinfo {author} {\bibfnamefont {C.-M.}\ \bibnamefont {Hu}},\ }\bibfield  {title} {\enquote {\bibinfo {title} {Coherent microwave emission of gain-driven polaritons},}\ }\href@noop {} {\bibfield  {journal} {\bibinfo  {journal} {Physical Review Letters}\ }\textbf {\bibinfo {volume} {130}},\ \bibinfo {pages} {146702} (\bibinfo {year} {2023})}\BibitemShut {NoStop}%
\bibitem [{\citenamefont {Zhang}\ \emph {et~al.}(2025)\citenamefont {Zhang}, \citenamefont {Kim}, \citenamefont {Zhang}, \citenamefont {Wang}, \citenamefont {Trivedi}, \citenamefont {Krasnok}, \citenamefont {Wang}, \citenamefont {Isleifson}, \citenamefont {Roshko},\ and\ \citenamefont {Hu}}]{zhang2025gain}%
  \BibitemOpen
  \bibfield  {author} {\bibinfo {author} {\bibfnamefont {C.}~\bibnamefont {Zhang}}, \bibinfo {author} {\bibfnamefont {M.}~\bibnamefont {Kim}}, \bibinfo {author} {\bibfnamefont {Y.-H.}\ \bibnamefont {Zhang}}, \bibinfo {author} {\bibfnamefont {Y.-P.}\ \bibnamefont {Wang}}, \bibinfo {author} {\bibfnamefont {D.}~\bibnamefont {Trivedi}}, \bibinfo {author} {\bibfnamefont {A.}~\bibnamefont {Krasnok}}, \bibinfo {author} {\bibfnamefont {J.}~\bibnamefont {Wang}}, \bibinfo {author} {\bibfnamefont {D.}~\bibnamefont {Isleifson}}, \bibinfo {author} {\bibfnamefont {R.}~\bibnamefont {Roshko}}, \ and\ \bibinfo {author} {\bibfnamefont {C.-M.}\ \bibnamefont {Hu}},\ }\bibfield  {title} {\enquote {\bibinfo {title} {Gain--loss coupled systems},}\ }\href@noop {} {\bibfield  {journal} {\bibinfo  {journal} {APL Quantum}\ }\textbf {\bibinfo {volume} {2}} (\bibinfo {year} {2025})}\BibitemShut {NoStop}%
\bibitem [{\citenamefont {Chembo}\ \emph {et~al.}(2019)\citenamefont {Chembo}, \citenamefont {Brunner}, \citenamefont {Jacquot},\ and\ \citenamefont {Larger}}]{chembo2019optoelectronic}%
  \BibitemOpen
  \bibfield  {author} {\bibinfo {author} {\bibfnamefont {Y.~K.}\ \bibnamefont {Chembo}}, \bibinfo {author} {\bibfnamefont {D.}~\bibnamefont {Brunner}}, \bibinfo {author} {\bibfnamefont {M.}~\bibnamefont {Jacquot}}, \ and\ \bibinfo {author} {\bibfnamefont {L.}~\bibnamefont {Larger}},\ }\bibfield  {title} {\enquote {\bibinfo {title} {Optoelectronic oscillators with time-delayed feedback},}\ }\href@noop {} {\bibfield  {journal} {\bibinfo  {journal} {Reviews of Modern Physics}\ }\textbf {\bibinfo {volume} {91}},\ \bibinfo {pages} {035006} (\bibinfo {year} {2019})}\BibitemShut {NoStop}%
\bibitem [{\citenamefont {Xiong}\ \emph {et~al.}(2024{\natexlab{c}})\citenamefont {Xiong}, \citenamefont {Nair}, \citenamefont {Christy}, \citenamefont {Cahoon}, \citenamefont {Pishehvar}, \citenamefont {Zhang}, \citenamefont {Flebus},\ and\ \citenamefont {Zhang}}]{xiong2024magnon}%
  \BibitemOpen
  \bibfield  {author} {\bibinfo {author} {\bibfnamefont {Y.}~\bibnamefont {Xiong}}, \bibinfo {author} {\bibfnamefont {J.~M.}\ \bibnamefont {Nair}}, \bibinfo {author} {\bibfnamefont {A.}~\bibnamefont {Christy}}, \bibinfo {author} {\bibfnamefont {J.~F.}\ \bibnamefont {Cahoon}}, \bibinfo {author} {\bibfnamefont {A.}~\bibnamefont {Pishehvar}}, \bibinfo {author} {\bibfnamefont {X.}~\bibnamefont {Zhang}}, \bibinfo {author} {\bibfnamefont {B.}~\bibnamefont {Flebus}}, \ and\ \bibinfo {author} {\bibfnamefont {W.}~\bibnamefont {Zhang}},\ }\bibfield  {title} {\enquote {\bibinfo {title} {Magnon-photon coupling in an opto-electro-magnonic oscillator},}\ }\href@noop {} {\bibfield  {journal} {\bibinfo  {journal} {npj Spintronics}\ }\textbf {\bibinfo {volume} {2}},\ \bibinfo {pages} {9} (\bibinfo {year} {2024}{\natexlab{c}})}\BibitemShut {NoStop}%
\bibitem [{\citenamefont {Dunn}\ and\ \citenamefont {Ebrahimzadeh}(1999)}]{dunn1999parametric}%
  \BibitemOpen
  \bibfield  {author} {\bibinfo {author} {\bibfnamefont {M.~H.}\ \bibnamefont {Dunn}}\ and\ \bibinfo {author} {\bibfnamefont {M.}~\bibnamefont {Ebrahimzadeh}},\ }\bibfield  {title} {\enquote {\bibinfo {title} {Parametric generation of tunable light from continuous-wave to femtosecond pulses},}\ }\href@noop {} {\bibfield  {journal} {\bibinfo  {journal} {Science}\ }\textbf {\bibinfo {volume} {286}},\ \bibinfo {pages} {1513--1517} (\bibinfo {year} {1999})}\BibitemShut {NoStop}%
\bibitem [{\citenamefont {Hao}\ \emph {et~al.}(2020)\citenamefont {Hao}, \citenamefont {Cen}, \citenamefont {Guan}, \citenamefont {Li}, \citenamefont {Dai}, \citenamefont {Zhu},\ and\ \citenamefont {Li}}]{hao2020optoelectronic}%
  \BibitemOpen
  \bibfield  {author} {\bibinfo {author} {\bibfnamefont {T.}~\bibnamefont {Hao}}, \bibinfo {author} {\bibfnamefont {Q.}~\bibnamefont {Cen}}, \bibinfo {author} {\bibfnamefont {S.}~\bibnamefont {Guan}}, \bibinfo {author} {\bibfnamefont {W.}~\bibnamefont {Li}}, \bibinfo {author} {\bibfnamefont {Y.}~\bibnamefont {Dai}}, \bibinfo {author} {\bibfnamefont {N.}~\bibnamefont {Zhu}}, \ and\ \bibinfo {author} {\bibfnamefont {M.}~\bibnamefont {Li}},\ }\bibfield  {title} {\enquote {\bibinfo {title} {Optoelectronic parametric oscillator},}\ }\href@noop {} {\bibfield  {journal} {\bibinfo  {journal} {Light: Science \& Applications}\ }\textbf {\bibinfo {volume} {9}},\ \bibinfo {pages} {102} (\bibinfo {year} {2020})}\BibitemShut {NoStop}%
\bibitem [{\citenamefont {Cen}\ \emph {et~al.}(2022)\citenamefont {Cen}, \citenamefont {Ding}, \citenamefont {Hao}, \citenamefont {Guan}, \citenamefont {Qin}, \citenamefont {Lyu}, \citenamefont {Li}, \citenamefont {Zhu}, \citenamefont {Xu}, \citenamefont {Dai} \emph {et~al.}}]{cen2022large}%
  \BibitemOpen
  \bibfield  {author} {\bibinfo {author} {\bibfnamefont {Q.}~\bibnamefont {Cen}}, \bibinfo {author} {\bibfnamefont {H.}~\bibnamefont {Ding}}, \bibinfo {author} {\bibfnamefont {T.}~\bibnamefont {Hao}}, \bibinfo {author} {\bibfnamefont {S.}~\bibnamefont {Guan}}, \bibinfo {author} {\bibfnamefont {Z.}~\bibnamefont {Qin}}, \bibinfo {author} {\bibfnamefont {J.}~\bibnamefont {Lyu}}, \bibinfo {author} {\bibfnamefont {W.}~\bibnamefont {Li}}, \bibinfo {author} {\bibfnamefont {N.}~\bibnamefont {Zhu}}, \bibinfo {author} {\bibfnamefont {K.}~\bibnamefont {Xu}}, \bibinfo {author} {\bibfnamefont {Y.}~\bibnamefont {Dai}},  \emph {et~al.},\ }\bibfield  {title} {\enquote {\bibinfo {title} {Large-scale coherent ising machine based on optoelectronic parametric oscillator},}\ }\href@noop {} {\bibfield  {journal} {\bibinfo  {journal} {Light: Science \& Applications}\ }\textbf {\bibinfo {volume} {11}},\ \bibinfo {pages} {333} (\bibinfo {year} {2022})}\BibitemShut {NoStop}%
\bibitem [{\citenamefont {Chembo}\ \emph {et~al.}(2008)\citenamefont {Chembo}, \citenamefont {Larger}, \citenamefont {Bendoula},\ and\ \citenamefont {Colet}}]{chembo2008effects}%
  \BibitemOpen
  \bibfield  {author} {\bibinfo {author} {\bibfnamefont {Y.~K.}\ \bibnamefont {Chembo}}, \bibinfo {author} {\bibfnamefont {L.}~\bibnamefont {Larger}}, \bibinfo {author} {\bibfnamefont {R.}~\bibnamefont {Bendoula}}, \ and\ \bibinfo {author} {\bibfnamefont {P.}~\bibnamefont {Colet}},\ }\bibfield  {title} {\enquote {\bibinfo {title} {Effects of gain and bandwidth on the multimode behavior of optoelectronic microwave oscillators},}\ }\href@noop {} {\bibfield  {journal} {\bibinfo  {journal} {Optics express}\ }\textbf {\bibinfo {volume} {16}},\ \bibinfo {pages} {9067--9072} (\bibinfo {year} {2008})}\BibitemShut {NoStop}%
\bibitem [{\citenamefont {Peil}\ \emph {et~al.}(2009)\citenamefont {Peil}, \citenamefont {Jacquot}, \citenamefont {Chembo}, \citenamefont {Larger},\ and\ \citenamefont {Erneux}}]{peil2009routes}%
  \BibitemOpen
  \bibfield  {author} {\bibinfo {author} {\bibfnamefont {M.}~\bibnamefont {Peil}}, \bibinfo {author} {\bibfnamefont {M.}~\bibnamefont {Jacquot}}, \bibinfo {author} {\bibfnamefont {Y.~K.}\ \bibnamefont {Chembo}}, \bibinfo {author} {\bibfnamefont {L.}~\bibnamefont {Larger}}, \ and\ \bibinfo {author} {\bibfnamefont {T.}~\bibnamefont {Erneux}},\ }\bibfield  {title} {\enquote {\bibinfo {title} {Routes to chaos and multiple time scale dynamics in broadband bandpass nonlinear delay electro-optic oscillators},}\ }\href@noop {} {\bibfield  {journal} {\bibinfo  {journal} {Physical Review E—Statistical, Nonlinear, and Soft Matter Physics}\ }\textbf {\bibinfo {volume} {79}},\ \bibinfo {pages} {026208} (\bibinfo {year} {2009})}\BibitemShut {NoStop}%
\bibitem [{\citenamefont {Inman}\ \emph {et~al.}(2022)\citenamefont {Inman}, \citenamefont {Xiong}, \citenamefont {Bidthanapally}, \citenamefont {Louis}, \citenamefont {Tyberkevych}, \citenamefont {Qu}, \citenamefont {Sklenar}, \citenamefont {Novosad}, \citenamefont {Li}, \citenamefont {Zhang} \emph {et~al.}}]{inman2022hybrid}%
  \BibitemOpen
  \bibfield  {author} {\bibinfo {author} {\bibfnamefont {J.}~\bibnamefont {Inman}}, \bibinfo {author} {\bibfnamefont {Y.}~\bibnamefont {Xiong}}, \bibinfo {author} {\bibfnamefont {R.}~\bibnamefont {Bidthanapally}}, \bibinfo {author} {\bibfnamefont {S.}~\bibnamefont {Louis}}, \bibinfo {author} {\bibfnamefont {V.}~\bibnamefont {Tyberkevych}}, \bibinfo {author} {\bibfnamefont {H.}~\bibnamefont {Qu}}, \bibinfo {author} {\bibfnamefont {J.}~\bibnamefont {Sklenar}}, \bibinfo {author} {\bibfnamefont {V.}~\bibnamefont {Novosad}}, \bibinfo {author} {\bibfnamefont {Y.}~\bibnamefont {Li}}, \bibinfo {author} {\bibfnamefont {X.}~\bibnamefont {Zhang}},  \emph {et~al.},\ }\bibfield  {title} {\enquote {\bibinfo {title} {Hybrid magnonics for short-wavelength spin waves facilitated by a magnetic heterostructure},}\ }\href@noop {} {\bibfield  {journal} {\bibinfo  {journal} {Physical Review Applied}\ }\textbf {\bibinfo {volume} {17}},\ \bibinfo {pages} {044034} (\bibinfo {year} {2022})}\BibitemShut {NoStop}%
\bibitem [{\citenamefont {Qu}\ \emph {et~al.}(2025)\citenamefont {Qu}, \citenamefont {Xiong}, \citenamefont {Zhang}, \citenamefont {Li},\ and\ \citenamefont {Zhang}}]{qu2025pump}%
  \BibitemOpen
  \bibfield  {author} {\bibinfo {author} {\bibfnamefont {T.}~\bibnamefont {Qu}}, \bibinfo {author} {\bibfnamefont {Y.}~\bibnamefont {Xiong}}, \bibinfo {author} {\bibfnamefont {X.}~\bibnamefont {Zhang}}, \bibinfo {author} {\bibfnamefont {Y.}~\bibnamefont {Li}}, \ and\ \bibinfo {author} {\bibfnamefont {W.}~\bibnamefont {Zhang}},\ }\bibfield  {title} {\enquote {\bibinfo {title} {Pump-induced magnon anticrossing due to three-magnon splitting and confluence},}\ }\href@noop {} {\bibfield  {journal} {\bibinfo  {journal} {Physical Review B}\ }\textbf {\bibinfo {volume} {111}},\ \bibinfo {pages} {L180410} (\bibinfo {year} {2025})}\BibitemShut {NoStop}%
\bibitem [{\citenamefont {Rao}\ \emph {et~al.}(2023{\natexlab{b}})\citenamefont {Rao}, \citenamefont {Yao}, \citenamefont {Wang}, \citenamefont {Zhang}, \citenamefont {Yu},\ and\ \citenamefont {Lu}}]{rao2023unveiling}%
  \BibitemOpen
  \bibfield  {author} {\bibinfo {author} {\bibfnamefont {J.}~\bibnamefont {Rao}}, \bibinfo {author} {\bibfnamefont {B.}~\bibnamefont {Yao}}, \bibinfo {author} {\bibfnamefont {C.}~\bibnamefont {Wang}}, \bibinfo {author} {\bibfnamefont {C.}~\bibnamefont {Zhang}}, \bibinfo {author} {\bibfnamefont {T.}~\bibnamefont {Yu}}, \ and\ \bibinfo {author} {\bibfnamefont {W.}~\bibnamefont {Lu}},\ }\bibfield  {title} {\enquote {\bibinfo {title} {Unveiling a pump-induced magnon mode via its strong interaction with walker modes},}\ }\href@noop {} {\bibfield  {journal} {\bibinfo  {journal} {Physical Review Letters}\ }\textbf {\bibinfo {volume} {130}},\ \bibinfo {pages} {046705} (\bibinfo {year} {2023}{\natexlab{b}})}\BibitemShut {NoStop}%
\bibitem [{\citenamefont {Rao}\ \emph {et~al.}(2025)\citenamefont {Rao}, \citenamefont {Wang}, \citenamefont {Chen}, \citenamefont {Yao}, \citenamefont {Zhao}, \citenamefont {Wei}, \citenamefont {Wang}, \citenamefont {Li}, \citenamefont {Bai},\ and\ \citenamefont {Lu}}]{rao2025time}%
  \BibitemOpen
  \bibfield  {author} {\bibinfo {author} {\bibfnamefont {J.}~\bibnamefont {Rao}}, \bibinfo {author} {\bibfnamefont {Y.-P.}\ \bibnamefont {Wang}}, \bibinfo {author} {\bibfnamefont {Z.}~\bibnamefont {Chen}}, \bibinfo {author} {\bibfnamefont {B.}~\bibnamefont {Yao}}, \bibinfo {author} {\bibfnamefont {K.}~\bibnamefont {Zhao}}, \bibinfo {author} {\bibfnamefont {C.}~\bibnamefont {Wei}}, \bibinfo {author} {\bibfnamefont {C.}~\bibnamefont {Wang}}, \bibinfo {author} {\bibfnamefont {R.}~\bibnamefont {Li}}, \bibinfo {author} {\bibfnamefont {L.}~\bibnamefont {Bai}}, \ and\ \bibinfo {author} {\bibfnamefont {W.}~\bibnamefont {Lu}},\ }\bibfield  {title} {\enquote {\bibinfo {title} {Time-varying strong coupling and the induced time diffraction of magnon modes},}\ }\href@noop {} {\bibfield  {journal} {\bibinfo  {journal} {Physical Review Letters}\ }\textbf {\bibinfo {volume} {135}},\ \bibinfo {pages} {066704} (\bibinfo {year} {2025})}\BibitemShut {NoStop}%
\bibitem [{\citenamefont {Bourhill}\ \emph {et~al.}(2016)\citenamefont {Bourhill}, \citenamefont {Kostylev}, \citenamefont {Goryachev}, \citenamefont {Creedon},\ and\ \citenamefont {Tobar}}]{bourhill2016ultrahigh}%
  \BibitemOpen
  \bibfield  {author} {\bibinfo {author} {\bibfnamefont {J.}~\bibnamefont {Bourhill}}, \bibinfo {author} {\bibfnamefont {N.}~\bibnamefont {Kostylev}}, \bibinfo {author} {\bibfnamefont {M.}~\bibnamefont {Goryachev}}, \bibinfo {author} {\bibfnamefont {D.}~\bibnamefont {Creedon}}, \ and\ \bibinfo {author} {\bibfnamefont {M.}~\bibnamefont {Tobar}},\ }\bibfield  {title} {\enquote {\bibinfo {title} {Ultrahigh cooperativity interactions between magnons and resonant photons in a yig sphere},}\ }\href@noop {} {\bibfield  {journal} {\bibinfo  {journal} {Physical Review B}\ }\textbf {\bibinfo {volume} {93}},\ \bibinfo {pages} {144420} (\bibinfo {year} {2016})}\BibitemShut {NoStop}%
\bibitem [{\citenamefont {Suhl}(1957)}]{Suhl57}%
  \BibitemOpen
  \bibfield  {author} {\bibinfo {author} {\bibfnamefont {H.}~\bibnamefont {Suhl}},\ }\bibfield  {title} {\enquote {\bibinfo {title} {{The theory of ferromagnetic resonance at high signal powers}},}\ }\href@noop {} {\bibfield  {journal} {\bibinfo  {journal} {J. Phys. Chem. Solids}\ }\textbf {\bibinfo {volume} {1}},\ \bibinfo {pages} {209} (\bibinfo {year} {1957})}\BibitemShut {NoStop}%
\bibitem [{\citenamefont {Pan}\ \emph {et~al.}(2024)\citenamefont {Pan}, \citenamefont {Cui}, \citenamefont {Yu}, \citenamefont {Zhang},\ and\ \citenamefont {Zhang}}]{pan2024frequency}%
  \BibitemOpen
  \bibfield  {author} {\bibinfo {author} {\bibfnamefont {Z.}~\bibnamefont {Pan}}, \bibinfo {author} {\bibfnamefont {S.}~\bibnamefont {Cui}}, \bibinfo {author} {\bibfnamefont {Y.}~\bibnamefont {Yu}}, \bibinfo {author} {\bibfnamefont {F.}~\bibnamefont {Zhang}}, \ and\ \bibinfo {author} {\bibfnamefont {X.}~\bibnamefont {Zhang}},\ }\bibfield  {title} {\enquote {\bibinfo {title} {Frequency-stabilized optoelectronic oscillator with simultaneous wideband tunability},}\ }\href@noop {} {\bibfield  {journal} {\bibinfo  {journal} {IEEE Transactions on Microwave Theory and Techniques}\ } (\bibinfo {year} {2024})}\BibitemShut {NoStop}%
\bibitem [{\citenamefont {Zhu}\ \emph {et~al.}(2020)\citenamefont {Zhu}, \citenamefont {Jin}, \citenamefont {Fu}, \citenamefont {Chi}, \citenamefont {Zuo}, \citenamefont {Liu},\ and\ \citenamefont {Wang}}]{zhu2020frequency}%
  \BibitemOpen
  \bibfield  {author} {\bibinfo {author} {\bibfnamefont {X.}~\bibnamefont {Zhu}}, \bibinfo {author} {\bibfnamefont {T.}~\bibnamefont {Jin}}, \bibinfo {author} {\bibfnamefont {Y.}~\bibnamefont {Fu}}, \bibinfo {author} {\bibfnamefont {H.}~\bibnamefont {Chi}}, \bibinfo {author} {\bibfnamefont {L.}~\bibnamefont {Zuo}}, \bibinfo {author} {\bibfnamefont {W.}~\bibnamefont {Liu}}, \ and\ \bibinfo {author} {\bibfnamefont {Q.}~\bibnamefont {Wang}},\ }\bibfield  {title} {\enquote {\bibinfo {title} {A frequency-stable optoelectronic oscillator based on passive phase compensation},}\ }\href@noop {} {\bibfield  {journal} {\bibinfo  {journal} {IEEE Photonics Technology Letters}\ }\textbf {\bibinfo {volume} {32}},\ \bibinfo {pages} {612--615} (\bibinfo {year} {2020})}\BibitemShut {NoStop}%
\bibitem [{\citenamefont {Zhang}, \citenamefont {Hou},\ and\ \citenamefont {Zhao}(2014)}]{zhang2014long}%
  \BibitemOpen
  \bibfield  {author} {\bibinfo {author} {\bibfnamefont {Y.}~\bibnamefont {Zhang}}, \bibinfo {author} {\bibfnamefont {D.}~\bibnamefont {Hou}}, \ and\ \bibinfo {author} {\bibfnamefont {J.}~\bibnamefont {Zhao}},\ }\bibfield  {title} {\enquote {\bibinfo {title} {Long-term frequency stabilization of an optoelectronic oscillator using phase-locked loop},}\ }\href@noop {} {\bibfield  {journal} {\bibinfo  {journal} {Journal of lightwave technology}\ }\textbf {\bibinfo {volume} {32}},\ \bibinfo {pages} {2408--2414} (\bibinfo {year} {2014})}\BibitemShut {NoStop}%
\bibitem [{\citenamefont {Valagiannopoulos}(2022)}]{valagiannopoulos2022multistability}%
  \BibitemOpen
  \bibfield  {author} {\bibinfo {author} {\bibfnamefont {C.}~\bibnamefont {Valagiannopoulos}},\ }\bibfield  {title} {\enquote {\bibinfo {title} {Multistability in coupled nonlinear metasurfaces},}\ }\href@noop {} {\bibfield  {journal} {\bibinfo  {journal} {IEEE Transactions on Antennas and Propagation}\ }\textbf {\bibinfo {volume} {70}},\ \bibinfo {pages} {5534--5540} (\bibinfo {year} {2022})}\BibitemShut {NoStop}%
\bibitem [{\citenamefont {Hula}\ \emph {et~al.}(2022)\citenamefont {Hula}, \citenamefont {Schultheiss}, \citenamefont {Goncalves}, \citenamefont {K{\"o}rber}, \citenamefont {Bejarano}, \citenamefont {Copus}, \citenamefont {Flacke}, \citenamefont {Liensberger}, \citenamefont {Buzdakov}, \citenamefont {K{\'a}kay} \emph {et~al.}}]{hula2022spin}%
  \BibitemOpen
  \bibfield  {author} {\bibinfo {author} {\bibfnamefont {T.}~\bibnamefont {Hula}}, \bibinfo {author} {\bibfnamefont {K.}~\bibnamefont {Schultheiss}}, \bibinfo {author} {\bibfnamefont {F.~J.~T.}\ \bibnamefont {Goncalves}}, \bibinfo {author} {\bibfnamefont {L.}~\bibnamefont {K{\"o}rber}}, \bibinfo {author} {\bibfnamefont {M.}~\bibnamefont {Bejarano}}, \bibinfo {author} {\bibfnamefont {M.}~\bibnamefont {Copus}}, \bibinfo {author} {\bibfnamefont {L.}~\bibnamefont {Flacke}}, \bibinfo {author} {\bibfnamefont {L.}~\bibnamefont {Liensberger}}, \bibinfo {author} {\bibfnamefont {A.}~\bibnamefont {Buzdakov}}, \bibinfo {author} {\bibfnamefont {A.}~\bibnamefont {K{\'a}kay}},  \emph {et~al.},\ }\bibfield  {title} {\enquote {\bibinfo {title} {Spin-wave frequency combs},}\ }\href@noop {} {\bibfield  {journal} {\bibinfo  {journal} {Applied Physics Letters}\ }\textbf {\bibinfo {volume} {121}} (\bibinfo {year} {2022})}\BibitemShut {NoStop}%
\bibitem [{\citenamefont {Wang}\ \emph {et~al.}(2024{\natexlab{b}})\citenamefont {Wang}, \citenamefont {Rao}, \citenamefont {Chen}, \citenamefont {Zhao}, \citenamefont {Sun}, \citenamefont {Yao}, \citenamefont {Yu}, \citenamefont {Wang},\ and\ \citenamefont {Lu}}]{wang2024enhancement}%
  \BibitemOpen
  \bibfield  {author} {\bibinfo {author} {\bibfnamefont {C.}~\bibnamefont {Wang}}, \bibinfo {author} {\bibfnamefont {J.}~\bibnamefont {Rao}}, \bibinfo {author} {\bibfnamefont {Z.}~\bibnamefont {Chen}}, \bibinfo {author} {\bibfnamefont {K.}~\bibnamefont {Zhao}}, \bibinfo {author} {\bibfnamefont {L.}~\bibnamefont {Sun}}, \bibinfo {author} {\bibfnamefont {B.}~\bibnamefont {Yao}}, \bibinfo {author} {\bibfnamefont {T.}~\bibnamefont {Yu}}, \bibinfo {author} {\bibfnamefont {Y.-P.}\ \bibnamefont {Wang}}, \ and\ \bibinfo {author} {\bibfnamefont {W.}~\bibnamefont {Lu}},\ }\bibfield  {title} {\enquote {\bibinfo {title} {Enhancement of magnonic frequency combs by exceptional points},}\ }\href@noop {} {\bibfield  {journal} {\bibinfo  {journal} {Nature Physics}\ }\textbf {\bibinfo {volume} {20}},\ \bibinfo {pages} {1139--1144} (\bibinfo {year} {2024}{\natexlab{b}})}\BibitemShut {NoStop}%
\bibitem [{\citenamefont {Christy}\ \emph {et~al.}(2025)\citenamefont {Christy}, \citenamefont {Zhu}, \citenamefont {Li}, \citenamefont {Xiong}, \citenamefont {Qu}, \citenamefont {Tsui}, \citenamefont {Cahoon}, \citenamefont {Yang}, \citenamefont {Hu},\ and\ \citenamefont {Zhang}}]{christy2025tuning}%
  \BibitemOpen
  \bibfield  {author} {\bibinfo {author} {\bibfnamefont {A.}~\bibnamefont {Christy}}, \bibinfo {author} {\bibfnamefont {Y.}~\bibnamefont {Zhu}}, \bibinfo {author} {\bibfnamefont {Y.}~\bibnamefont {Li}}, \bibinfo {author} {\bibfnamefont {Y.}~\bibnamefont {Xiong}}, \bibinfo {author} {\bibfnamefont {T.}~\bibnamefont {Qu}}, \bibinfo {author} {\bibfnamefont {F.}~\bibnamefont {Tsui}}, \bibinfo {author} {\bibfnamefont {J.~F.}\ \bibnamefont {Cahoon}}, \bibinfo {author} {\bibfnamefont {B.}~\bibnamefont {Yang}}, \bibinfo {author} {\bibfnamefont {J.-M.}\ \bibnamefont {Hu}}, \ and\ \bibinfo {author} {\bibfnamefont {W.}~\bibnamefont {Zhang}},\ }\bibfield  {title} {\enquote {\bibinfo {title} {Tuning magneto-optical zero reflection via dual-channel hybrid magnonics},}\ }\href@noop {} {\bibfield  {journal} {\bibinfo  {journal} {Physical Review Applied}\ }\textbf {\bibinfo {volume} {24}},\ \bibinfo {pages} {034045} (\bibinfo {year} {2025})}\BibitemShut {NoStop}%
\bibitem [{\citenamefont {Xu}\ \emph {et~al.}(2024{\natexlab{b}})\citenamefont {Xu}, \citenamefont {Cheung}, \citenamefont {Cormode}, \citenamefont {Puel}, \citenamefont {Pal}, \citenamefont {Yusuf}, \citenamefont {Chilcote}, \citenamefont {Flatt{\'e}}, \citenamefont {Johnston-Halperin},\ and\ \citenamefont {Fuchs}}]{xu2024strong}%
  \BibitemOpen
  \bibfield  {author} {\bibinfo {author} {\bibfnamefont {Q.}~\bibnamefont {Xu}}, \bibinfo {author} {\bibfnamefont {H.~F.~H.}\ \bibnamefont {Cheung}}, \bibinfo {author} {\bibfnamefont {D.~S.}\ \bibnamefont {Cormode}}, \bibinfo {author} {\bibfnamefont {T.~O.}\ \bibnamefont {Puel}}, \bibinfo {author} {\bibfnamefont {S.}~\bibnamefont {Pal}}, \bibinfo {author} {\bibfnamefont {H.}~\bibnamefont {Yusuf}}, \bibinfo {author} {\bibfnamefont {M.}~\bibnamefont {Chilcote}}, \bibinfo {author} {\bibfnamefont {M.~E.}\ \bibnamefont {Flatt{\'e}}}, \bibinfo {author} {\bibfnamefont {E.}~\bibnamefont {Johnston-Halperin}}, \ and\ \bibinfo {author} {\bibfnamefont {G.~D.}\ \bibnamefont {Fuchs}},\ }\bibfield  {title} {\enquote {\bibinfo {title} {Strong photon-magnon coupling using a lithographically defined organic ferrimagnet},}\ }\href@noop {} {\bibfield  {journal} {\bibinfo  {journal} {Advanced science}\ }\textbf {\bibinfo {volume} {11}},\ \bibinfo {pages} {2310032} (\bibinfo {year} {2024}{\natexlab{b}})}\BibitemShut {NoStop}%
\bibitem [{\citenamefont {Joseph}\ \emph {et~al.}(2024)\citenamefont {Joseph}, \citenamefont {Nair}, \citenamefont {Smith}, \citenamefont {Holland}, \citenamefont {McLellan}, \citenamefont {Boventer}, \citenamefont {Wolz}, \citenamefont {Bozhko}, \citenamefont {Flebus}, \citenamefont {Weides} \emph {et~al.}}]{joseph2024role}%
  \BibitemOpen
  \bibfield  {author} {\bibinfo {author} {\bibfnamefont {A.}~\bibnamefont {Joseph}}, \bibinfo {author} {\bibfnamefont {J.~M.}\ \bibnamefont {Nair}}, \bibinfo {author} {\bibfnamefont {M.~A.}\ \bibnamefont {Smith}}, \bibinfo {author} {\bibfnamefont {R.}~\bibnamefont {Holland}}, \bibinfo {author} {\bibfnamefont {L.~J.}\ \bibnamefont {McLellan}}, \bibinfo {author} {\bibfnamefont {I.}~\bibnamefont {Boventer}}, \bibinfo {author} {\bibfnamefont {T.}~\bibnamefont {Wolz}}, \bibinfo {author} {\bibfnamefont {D.~A.}\ \bibnamefont {Bozhko}}, \bibinfo {author} {\bibfnamefont {B.}~\bibnamefont {Flebus}}, \bibinfo {author} {\bibfnamefont {M.~P.}\ \bibnamefont {Weides}},  \emph {et~al.},\ }\bibfield  {title} {\enquote {\bibinfo {title} {The role of excitation vector fields and all-polarisation state control in cavity magnonics},}\ }\href@noop {} {\bibfield  {journal} {\bibinfo  {journal} {npj Spintronics}\ }\textbf {\bibinfo {volume} {2}},\ \bibinfo {pages} {59} (\bibinfo {year} {2024})}\BibitemShut {NoStop}%
\bibitem [{\citenamefont {Gardin}\ \emph {et~al.}(2023)\citenamefont {Gardin}, \citenamefont {Bourhill}, \citenamefont {Vlaminck}, \citenamefont {Person}, \citenamefont {Fumeaux}, \citenamefont {Castel},\ and\ \citenamefont {Tettamanzi}}]{gardin2023manifestation}%
  \BibitemOpen
  \bibfield  {author} {\bibinfo {author} {\bibfnamefont {A.}~\bibnamefont {Gardin}}, \bibinfo {author} {\bibfnamefont {J.}~\bibnamefont {Bourhill}}, \bibinfo {author} {\bibfnamefont {V.}~\bibnamefont {Vlaminck}}, \bibinfo {author} {\bibfnamefont {C.}~\bibnamefont {Person}}, \bibinfo {author} {\bibfnamefont {C.}~\bibnamefont {Fumeaux}}, \bibinfo {author} {\bibfnamefont {V.}~\bibnamefont {Castel}}, \ and\ \bibinfo {author} {\bibfnamefont {G.~C.}\ \bibnamefont {Tettamanzi}},\ }\bibfield  {title} {\enquote {\bibinfo {title} {Manifestation of the coupling phase in microwave cavity magnonics},}\ }\href@noop {} {\bibfield  {journal} {\bibinfo  {journal} {Physical Review Applied}\ }\textbf {\bibinfo {volume} {19}},\ \bibinfo {pages} {054069} (\bibinfo {year} {2023})}\BibitemShut {NoStop}%
\bibitem [{\citenamefont {Li}\ \emph {et~al.}(2019)\citenamefont {Li}, \citenamefont {Polakovic}, \citenamefont {Wang}, \citenamefont {Xu}, \citenamefont {Lendinez}, \citenamefont {Zhang}, \citenamefont {Ding}, \citenamefont {Khaire}, \citenamefont {Saglam}, \citenamefont {Divan} \emph {et~al.}}]{li2019strong}%
  \BibitemOpen
  \bibfield  {author} {\bibinfo {author} {\bibfnamefont {Y.}~\bibnamefont {Li}}, \bibinfo {author} {\bibfnamefont {T.}~\bibnamefont {Polakovic}}, \bibinfo {author} {\bibfnamefont {Y.-L.}\ \bibnamefont {Wang}}, \bibinfo {author} {\bibfnamefont {J.}~\bibnamefont {Xu}}, \bibinfo {author} {\bibfnamefont {S.}~\bibnamefont {Lendinez}}, \bibinfo {author} {\bibfnamefont {Z.}~\bibnamefont {Zhang}}, \bibinfo {author} {\bibfnamefont {J.}~\bibnamefont {Ding}}, \bibinfo {author} {\bibfnamefont {T.}~\bibnamefont {Khaire}}, \bibinfo {author} {\bibfnamefont {H.}~\bibnamefont {Saglam}}, \bibinfo {author} {\bibfnamefont {R.}~\bibnamefont {Divan}},  \emph {et~al.},\ }\bibfield  {title} {\enquote {\bibinfo {title} {Strong coupling between magnons and microwave photons in on-chip ferromagnet-superconductor thin-film devices},}\ }\href@noop {} {\bibfield  {journal} {\bibinfo  {journal} {Physical review letters}\ }\textbf {\bibinfo {volume} {123}},\ \bibinfo {pages} {107701} (\bibinfo {year} {2019})}\BibitemShut {NoStop}%
\bibitem [{\citenamefont {Hou}\ and\ \citenamefont {Liu}(2019)}]{hou2019strong}%
  \BibitemOpen
  \bibfield  {author} {\bibinfo {author} {\bibfnamefont {J.~T.}\ \bibnamefont {Hou}}\ and\ \bibinfo {author} {\bibfnamefont {L.}~\bibnamefont {Liu}},\ }\bibfield  {title} {\enquote {\bibinfo {title} {Strong coupling between microwave photons and nanomagnet magnons},}\ }\href@noop {} {\bibfield  {journal} {\bibinfo  {journal} {Physical review letters}\ }\textbf {\bibinfo {volume} {123}},\ \bibinfo {pages} {107702} (\bibinfo {year} {2019})}\BibitemShut {NoStop}%
\end{thebibliography}%

\end{document}